\documentclass[%
 aip,
 amsmath,amssymb,
reprint, 
]{revtex4-1}

\usepackage{color}
\usepackage{graphicx}
\usepackage{dcolumn}
\usepackage{bm}

\usepackage[utf8]{inputenc}
\usepackage[T1]{fontenc}
\usepackage{mathptmx}
\usepackage{etoolbox}

\def\aaa{{\it a}}
\def\bbb{{\it b}}
\def\ccc{{\it c}}
\def\ddd{{\it d}}
\def\eee{{\it e}}
\def\fff{{\it f}}
\newcommand{\mylab}[3]{\raisebox{#2}[0mm][0mm]{%
\makebox[0mm][l]{\hspace*{#1}\textbf{#3}}}}

\makeatletter
\def\@email#1#2{%
 \endgroup
 \patchcmd{\titleblock@produce}
  {\frontmatter@RRAPformat}
  {\frontmatter@RRAPformat{\produce@RRAP{*#1\href{mailto:#2}{#2}}}\frontmatter@RRAPformat}
  {}{}
}%
\makeatother
\begin{document}

\preprint{AIP/123-QED}

\title[Unifying formalism and closures for coarse-grained approaches to turbulence]{Unifying formalism and closures for coarse-grained approaches to turbulence}
\author{A. Cimarelli}
\email{andrea.cimarelli@unimore.it}
\affiliation{Department of Engineering "Enzo Ferrari", University of Modena and Reggio Emilia, 41125 Modena, Italy}
\author{N. Marras}
\affiliation{Department of Engineering "Enzo Ferrari", University of Modena and Reggio Emilia, 41125 Modena, Italy}
\author{B. Niceno}
\affiliation{Paul Scherrer Institute, ETH, 5232 Villigen, Switzerland}
\author{Y. Tessier Urrecha}
\affiliation{Department of Engineering "Enzo Ferrari", University of Modena and Reggio Emilia, 41125 Modena, Italy}

\date{\today}

\begin{abstract}
We propose the use of an unifying paradigm for the assessment and development of closed forms of the  coarse-grained Navier-Stokes equations in approaches ranging from the statistical to the scale-resolving ones. It consists in the exact formalism provided by the temporally filtered Navier-Stokes equations. The fundamental idea is that the smoothing action of turbulent stresses can be described as a temporal filtering operator implicitly applied to the solution. Contrary to the average and spatial filtering operators, the temporal filter is an unifying operator smoothly varying within the statistical and scale-resolving realms. The potential of the temporal filtering paradigm is here highlighted by unveiling relevant algebraic properties and by deriving a new class of turbulence closures. A dynamic procedure is derived and shown to provide an unifying closure for both scale-resolving and statistical approaches. Results show that an improved physics is captured. Challenging phenomena such as the laminar to turbulence transition and the dependence of separation and reattachment on free-stream turbulence applied through boundary conditions are captured.
\end{abstract}

\maketitle


\section{Introduction}
\label{sec_intro}

The more classical coarse-grained approach to the Navier-Stokes equations is that provided by the Reynolds-averaged Navier-Stokes equations (RANS). In their derivation an ensemble average operator $\langle \cdot \rangle$ is adopted. Associated with this operator is the appearance in the momentum equations of the Reynolds stresses which, accordingly, can be understood as second-order statistical moments of the velocity field,
\begin{equation}
R(u_i,u_j) = \langle u_i u_j \rangle - \langle u_i \rangle \langle u_j \rangle
\end{equation}
Because of the Reynolds stresses, the system of equations is however not closed thus posing the question on how to close the system \citep{wilcox1998turbulence}. The most classical way is to consider a gradient hypothesis such that the turbulent transport of mean momentum goes down the mean velocity gradient, the so-called Boussinesq hypothesis of eddy viscosity. Clearly, the solution is not anymore the solution of the exact RANS equations but of a different set of equations that depends on the model used for the turbulent stresses. This set of equations will be hereafter called modeled RANS equations. Accordingly, their solution cannot be anymore interpreted as the ensemble average solution of the Navier-Stokes equations. As a matter of facts, it is well known that classical turbulence closures predict unsteady solutions in flow problems with a large separation of scales whose ensemble solutions are on the contrary steady. Depending on the additional transport of momentum introduced by the turbulence closure, the solution gets different statistical features that are not present in the ensemble average. Thus, one might conclude that the solutions of the modeled equations are simply wrong. However, an alternative interpretation could be equivalently valid. Indeed, it is also possible to think that this discrepancy in the solutions is a symptom of the fact that the modeled equations represent a coarse-grained approach whose paradigm is different from that provided by the exact RANS equations. In other words, the reference for the solutions of the modeled RANS equations cannot be anymore the exact RANS solution and any discrepancies do not necessarily lead to the conclusion that the solutions of the modeled equations are wrong. If that is the case, the question is: is it possible to define a general formalism able to characterize the different types of coarse-grained approaches established by the different turbulence closures? Can this formalism be used to improve turbulence closures? These questions are even more relevant when considering hybrid turbulence closures \citep{menter2005scale, frohlich2008hybrid, spalart2009detached, chaouat2017state, heinz2020review, menter2021overview} where scale-resolving models typical of Large Eddy Simulation (LES) \citep{lesieur1996new, piomelli1999large, meneveau2000scale, moser2021statistical} are combined with statistical models typical of RANS \citep{launder1983numerical, spalart1992one, menter1994two, wilcox2008formulation, SSTDES}. The resulting coarse-grained approach is an ambiguous combination of statistical and scale-resolving methods that again lacks a general exact formalism able to describe them simultaneously.

In the present work, we propose the use of an unifying paradigm able to simultaneously describe both statistical and scale-resolving methods for the coarse-grained solution of the Navier-Stokes equations. The potential of the proposed formalism --which has been addressed in works such as \citet{pruett2003temporally,FadaiGhotbi}-- is demonstrated in the present work with the derivation of a new class of dynamic turbulence closures able to address solution features typical of both scale-resolving and statistical methods.

The paper is organized as follows. The theoretical framework is introduced in section \S\ref{sec_theory}. A family of improved turbulence closures is described in section \S\ref{sec_dyn}. Tests are performed for a transitional flow over a rounded leading-edge flat plate. The numerical settings and the flow setup are outlined in section \S\ref{sec_flowsetup} while the results from scale-resolving and statistical simulations are reported in sections \S\ref{sec_3D_results} and \S\ref{sec_2D_results}, respectively. The paper is closed by concluding remarks in section \S\ref{sec_concl}, by two appendices further assessing the improved turbulence closures, sections \S\ref{app_barc} and \S\ref{app_bump}, and by other two appendices addressing the effect of specific settings of the model and its computational costs, sections \S\ref{app_effects} and \S\ref{app_comp_costs}.


\section{Theoretical framework}
\label{sec_theory}

The funding aspect of the present work is the use of an operatorial approach for the description of the smoothing action given by the additional momentum transport introduced by turbulence models commonly employed in RANS and hybrid LES/RANS approaches. The basic assumption is that the smoothing action of the turbulence closure can be modeled as a temporal filter operator implicitly applied to the Navier-Stokes equations. This operator is hereafter denoted as $\langle \cdot \rangle_\tau$, where $\tau$ represents the filter time scale implicitly imposed by the turbulence closure. This operatorial assumption suggests that the modeled equations commonly solved in RANS and hybrid LES/RANS approaches can be better studied within the exact framework provided by the temporally filtered Navier-Stokes equations,
\begin{equation}
\left .
\begin{split}
&\frac{\partial \langle u_i \rangle_\tau}{\partial x_i} = 0 \\
&\frac{\partial \langle u_i \rangle_\tau}{\partial t} + \frac{\partial \langle u_i  \rangle_\tau \langle u_j \rangle_\tau}{\partial x_j} = -\frac{1}{\rho} \frac{\partial \langle p \rangle_\tau}{\partial x_i} + \nu \frac{\partial^2 \langle u_i \rangle_\tau}{\partial x_j\partial x_j} - \frac{\partial R_\tau(u_i,u_j)}{\partial x_j}
\end{split}
\right \}
\label{eq_filter}
\end{equation}
where $R_\tau(u_i,u_j)$ represents the tensor of turbulent stresses that in the modeled RANS and LES/RANS equations is provided by the turbulence closure.

Before addressing the closure issue and its relation with the temporal filtering operator, it is important to address first the theoretical framework provided by the temporally filtered Navier-Stokes equations in their exact form. To note that this framework has already attracted some interest in the scientific community. The reader is referred to the works \citet{pruett2003temporally}, \citet{pruett2008temporal} and references therein for a detailed assessment of the temporally filtered framework. Here, we report the main properties that are of relevance for the present work. In the exact framework provided by the temporally filtered Navier-Stokes equations (\ref{eq_filter}), the turbulent stress tensor takes the exact form,
\begin{equation}
R_\tau (u_i,u_j) = \langle u_i u_j \rangle_\tau - \langle u_i \rangle_\tau \langle u_j \rangle_\tau
\label{eq_exact_temp_stress}
\end{equation}
and the time filtering operator is not implicitly induced by the smoothing action of the turbulence closure, rather is explicitly applied. For a generic quantity $\beta$, the explicit filter operator can be formally and quite generally expressed as a convolution integral \citep{leonard1975energy},
\begin{equation}
    \langle \beta (\bm{x},t)\rangle_\tau = \int \beta (\bm{x},t') G (t-t', \tau) \, dt'
\end{equation}
where $G$ is a filter kernel with characteristic time scale $\tau$ that we assume to be constant preserving,
\begin{equation}
    \int G (t-t', \tau) \, dt' = 1
\end{equation}
and we assume to be time invariant to allow its commutation with temporal derivatives. The reader is referred to section 5.4 of \citet{moser2021statistical} for a review of works dedicated to the commutation errors, and to \citet{leonard2007commutation} and \citet{franke2001temporal} specifically for temporal commutation errors.

As already shown by previous pivotal works like those carried out by \citet{fadai2010temporal}, \citet{friess2015toward} and \citet{duffal2022development}, a peculiar and relevant aspect of the exact formalism provided by the temporally filtered Navier-Stokes equations is the direct link between the scale-resolving and the statistical approaches. Indeed, from a point of view of operators we have that the ensemble average is asymptotically recovered by the temporal filter $\langle \cdot \rangle_\tau = \langle \cdot \rangle$ when $\tau U/L \gg 1$ with $U$ and $L$ the characteristic velocity and length scales of the problem, provided that the flow has a statistically steady solution \citep{kampe1951theoretical}. This asymptotic link between the scale-resolving and statistical approaches is recovered also by the solution of the temporally filtered Navier-Stokes equations (\ref{eq_filter}). Indeed, when $\tau U/L \gg 1$ the exact turbulent stresses reduce to the exact Reynolds stresses, $R_\tau (u_i,u_j) = R(u_i,u_j)$ because $\langle u_i u_j \rangle_\tau - \langle u_i \rangle_\tau \langle u_j \rangle_\tau = \langle u_i u_j \rangle - \langle u_i \rangle \langle u_j \rangle$, hence the solution of the temporally filtered Navier-Stokes equations (\ref{eq_filter}) converges to the solution of the exact RANS equations, namely $\langle u_i \rangle_\tau = \langle u_i \rangle$. In other words, the temporally filtered Navier-Stokes equations represent an unifying theoretical framework able to describe the behaviour of solutions from turbulence closures ranging from the scale-resolving to the statistical approach. This property is missed by the exact spatially filtered Navier-Stokes equations and by the exact RANS equations that can describe only the scale-resolving and the statistical approaches separately. The aim of the present work is to support this change of paradigm by showing how and to what degree the temporal filtering formalism can be used to assess and eventually improve turbulence closures commonly adopted to solve the modeled RANS and hybrid LES/RANS equations.

A relevant aspect of the use of the temporal filtering approach is given by the fact that some important algebraic properties can be leveraged by considering different filtering operators hierarchically organized at multiple levels \citep{germano1992turbulence}. In particular, by introducing a second filter operator with a larger characteristic time scale $T> \tau$, it is possible to evaluate the turbulent stresses at a coarser level as
\begin{equation}
R_{T_\tau}(u_i,u_j) =\langle \langle u_i u_j \rangle_\tau \rangle_T - \langle \langle u_i \rangle_\tau \rangle_T \langle \langle u_j \rangle_\tau \rangle_T
\end{equation}
These turbulent stresses are related with those at the finer level as
\begin{equation}
R_{T_\tau}(u_i,u_j) = \langle R_\tau(u_i,u_j) \rangle_T + \langle \langle u_i \rangle_\tau \langle u_j \rangle_\tau \rangle_T - \langle \langle u_i \rangle_\tau \rangle_T \langle \langle u_j \rangle_\tau \rangle_T
\end{equation}
thus allowing us to write the following exact equation
\begin{equation}
R_{T_\tau}(u_i,u_j) - \langle R_\tau(u_i,u_j) \rangle_T = \langle \langle u_i \rangle_\tau \langle u_j \rangle_\tau \rangle_T - \langle \langle u_i \rangle_\tau \rangle_T \langle \langle u_j \rangle_\tau \rangle_T
\label{identity}
\end{equation}
which is the temporal analogue of one of the most famous equations in turbulence -- the Germano identity \citep{germano1992turbulence}. This exact algebraic relation is exploited here for the first time to derive a general approach for improving turbulence closures commonly adopted in RANS and hybrid LES/RANS approaches.


\section{Improved turbulence closures}
\label{sec_dyn}

The basic assumption of the present work is that the smoothing action of the additional momentum transport induced by turbulence closures can be modeled as an implicit filtering of the flow solution at a time scale $\tau$ that, without loss of generality, can be estimated as
\begin{equation}
\tau = \sqrt{\frac{\nu+\nu_\tau}{\epsilon}}
\end{equation}
where $\nu_\tau$ is the eddy viscosity associated with the turbulence closure while $\epsilon$ is the dissipation rate of kinetic energy. Accordingly, the solution from turbulence closures is expected to reduce to a statistically steady state when $\tau U/L \gg 1$. In this case, the reference framework is that provided by the exact RANS equations or equivalently by the large time scale limit of the temporally filtered Navier-Stokes equations. On the other hand, for flow cases characterized by a large separation of scales, the solution from turbulence closures is usually able to capture some unsteadiness since $\tau U/L \le 1$. In this condition, the paradigm for turbulence closures is no longer that provided by the exact RANS equations, rather only the exact framework provided by the temporally filtered Navier-Stokes equations can be used. Hence, this change of paradigm describes both the statistical and scale-resolving features of the solutions from turbulence closures and opens the possibility of improving them. The use of identity (\ref{identity}) and of the related decomposition of the velocity field at two filter levels $\tau$ and $T$, is a sound starting point to reach this goal.

\subsection{Dynamic RANS closure}
\label{dynev}

In the context of turbulence closures and, hence, of the modeled temporally filtered Navier-Stokes equations, the decomposition of the solution at two filter levels suggested by the identity (\ref{identity}) takes the following interpretation. The first filter operator at the time scale $\tau$ should be understood as implicitly applied by the eddy viscosity of the closure itself. Hence, all the quantities hereafter formally reported under the first filter operator $\langle \cdot \rangle_\tau$ actually represent quantities that are simply resolved. On the contrary, the second filter operator $\langle \cdot \rangle_T$ is the only explicitly filtering operation performed at a time scale $T> \tau$.

The formally exact algebraic relation (\ref{identity}), requires to model turbulent stresses at the resolved and test filter level, $R_\tau(u_i,u_j)$ and $R_{T_\tau}(u_i,u_j)$ respectively. The classical approach to model the turbulent stresses is to consider the Boussinesq gradient hypothesis by which the momentum is assumed to flow down the resolved velocity gradient. By applying this assumption we can write
\begin{equation}
\begin{split}
&R_\tau(u_i,u_j) = -2 \nu_\tau \langle S_{ij} \rangle_\tau \\
&R_{T_\tau}(u_i,u_j) = -2 \nu_{T_\tau} \langle \langle S_{ij} \rangle_\tau \rangle_T
\end{split}
\label{eq_boussinesq}
\end{equation}
where $S_{ij} = (\partial u_i/\partial x_j + \partial u_j/\partial x_i) /2$ is the rate of deformation tensor and the two diffusion coefficients, $\nu_\tau$ and $\nu_{T_\tau}$, are the eddy viscosity associated to the two filter levels. The latter are commonly modeled through turbulent quantities obtained from dedicated evolution equations \citep{wilcox1998turbulence}. Without loss of generality, let us consider as transported turbulent quantities, the turbulent kinetic energy and dissipation of the decomposed field, 
\begin{equation}
\begin{split}
&\nu_\tau = c_\mu \frac{k_\tau^2}{\epsilon_\tau} \\
&\nu_{T_\tau} = c_\mu \frac{k_{T_\tau}^2}{\epsilon_{T_\tau}}
\end{split}
\label{eq_k_eps}
\end{equation}
where $c_\mu$ is usually a constant coefficient and is assumed here to be the same for the two solutions at the two filter levels \citep{meneveau2000scale}. Here, $k_\tau$ and $\epsilon_\tau$ are the turbulent kinetic energy and turbulent dissipation associated with the resolved flow solution $\langle \bm{u} \rangle_\tau$ and are computed through their transport equations. Again without loss of generality, such equations can be of the following form,
\begin{equation}
\begin{split}
&\frac{\partial k_\tau}{\partial t} + \frac{\partial k_\tau \langle u_j \rangle_\tau}{\partial x_j} = \\
&2 \nu_\tau \langle S_{ij} \rangle_\tau \langle S_{ij} \rangle_\tau  - \epsilon_\tau + \frac{\partial}{\partial x_j} \left[ 
 \left ( \nu + \frac{\nu_\tau}{\sigma^k} \right) \frac{\partial k_\tau}{\partial x_j} \right] \\
\end{split}
\end{equation}
\begin{equation}
\begin{split}
&\frac{\partial \epsilon_\tau}{\partial t} + \frac{\partial \epsilon_\tau \langle u_j \rangle_\tau}{\partial x_j} = \\
&\frac{\epsilon_\tau}{k_\tau} \bigg ( C_{\epsilon_1}2 \nu_\tau \langle S_{ij} \rangle_\tau \langle S_{ij} \rangle_\tau  - C_{\epsilon_2} f_\epsilon \epsilon_\tau \bigg ) + \frac{\partial}{\partial x_j} \left[ 
 \left ( \nu + \frac{\nu_\tau}{\sigma^\epsilon} \right) \frac{\partial \epsilon_\tau}{\partial x_j} \right]
\end{split}
\label{eq_eps}
\end{equation}
where $\sigma^k$, $\sigma^\epsilon$, $C_{\epsilon_1}$ and $C_{\epsilon_2}$ are adjustable model coefficients \citep{launder1983numerical} and $f_\epsilon$ is a near-wall damping function to correct the near-wall behaviour \citep{abe1995new}. On the other hand, $k_{T_\tau}$ and $\epsilon_{T_\tau}$ are the turbulent kinetic energy and turbulent dissipation associated with the flow solution at the second filter level $\langle \langle \bm{u} \rangle_\tau \rangle_T$. These latter are not transported, rather can be computed as
\begin{equation}
\begin{split}
&k_{T_\tau} = \langle k_\tau \rangle_T + \mathcal{L}_{ii} / 2\\
&\epsilon_{T_\tau} = \langle \epsilon_\tau \rangle_T + \nu \mathcal{D}_{ii}
\end{split}
\label{eq_estimate_k_e}
\end{equation}
where the tensors $\mathcal{L}_{ij}$ and $\mathcal{D}_{ij}$ take origin from the same algebraic property introduced by the use of filtering formalism when comparing hierarchically organized filter levels. In particular, $\mathcal{L}_{ij}$ directly comes from identity (\ref{identity}),
\begin{equation}
\begin{split}
\mathcal{L}_{ij} \equiv R_{T_\tau}(u_i,u_j) - \langle R_\tau(u_i,u_j) \rangle_T = \\
\langle \langle u_i \rangle_\tau \langle u_j \rangle_\tau \rangle_T - \langle \langle u_i \rangle_\tau \rangle_T \langle \langle u_j \rangle_\tau \rangle_T
\end{split}
\end{equation}
on the other hand, $\mathcal{D}_{ij}$ derives from the same arguments but applied to the velocity gradient,
\begin{equation}
\begin{split}
\mathcal{D}_{ij} \equiv &R_{T_\tau} \left (\frac{\partial u_i}{\partial x_k},\frac{\partial u_j}{\partial x_k} \right ) - \left \langle R_\tau\left (\frac{\partial u_i}{\partial x_k},\frac{\partial u_j}{\partial x_k} \right ) \right \rangle_T = \\
& \left \langle \frac{\partial \langle u_i \rangle_\tau}{\partial x_k} \frac{\partial \langle u_j \rangle_\tau}{\partial x_k} \right \rangle_T - \frac{\partial \langle \langle u_i \rangle_\tau \rangle_T}{\partial x_k} \frac{\partial \langle \langle u_j \rangle_\tau \rangle_T}{\partial x_k}
\end{split}
\end{equation}

All the arguments are now in place to exploit identity (\ref{identity}) for the development of a generic framework for the improvement of turbulence closures. By substituting modeling assumptions (\ref{eq_k_eps}) into the Boussinesq hypothesis (\ref{eq_boussinesq}), the identity (\ref{identity}) can be rewritten as
\begin{equation}
\begin{split}
2 c_\mu \left \langle \frac{k_\tau^2}{\epsilon_\tau} \langle S_{ij} \rangle_\tau \right \rangle_T  -2 c_\mu  \frac{k_{T_\tau}^2}{\epsilon_{T_\tau}} \langle \langle S_{ij} \rangle_\tau \rangle_T = \\
\langle \langle u_i \rangle_\tau \langle u_j \rangle_\tau \rangle_T - \langle \langle u_i \rangle_\tau \rangle_T \langle \langle u_j \rangle_\tau \rangle_T
\end{split}
\label{eq_dyn_eq}
\end{equation}
that in a more compact form reads
\begin{equation}
c_\mu \, \mathcal{M}_{ij} = \mathcal{L}_{ij}
\label{eq_dyn_eq_compact}
\end{equation}
where
\begin{equation}
\mathcal{M}_{ij} = 2 \left \langle \frac{k_\tau^2}{\epsilon_\tau} \langle S_{ij} \rangle_\tau \right \rangle_T  -2 \frac{k_{T_\tau}^2}{\epsilon_{T_\tau}} \langle \langle S_{ij} \rangle_\tau \rangle_T
\end{equation}
In equations (\ref{eq_dyn_eq_compact}) the only unknown is the model coefficient $c_\mu$ that can be easily computed as the value that minimizes the mean square error of the system \citep{lilly1992proposed},
\begin{equation}
c_\mu = \frac{\mathcal{M}_{ij} \mathcal{L}_{ij}}{\mathcal{M}_{ij} \mathcal{M}_{ij}}
\label{eq_dyn_eq}
\end{equation}
Equation (\ref{eq_dyn_eq}) represents a new class of turbulence closures. Indeed, the tensor $\mathcal{M}_{ij}$ has been derived here specifically for the $k-\varepsilon$ model but can be equivalently written for any other two-equation eddy viscosity model. This new class of models (\ref{eq_dyn_eq}) will be hereafter called dynamic RANS closures. To note that this procedure is similar to and directly inspired by that pursued in Large-Eddy Simulation (LES) --to derive the dynamic Smagorinsky model \citep{germano1991dynamic}--  and in hybrid approaches such as Detached eddy simulation (DES) \citep{yin2015dynamic}. The main difference is given by the fact that the new dynamic RANS approach is based on the temporal filtering formalism and aims at improving the predictions of both statistical and scale-resolving approaches typical of RANS and hybrid LES/RANS solutions.

\subsection{Foreseen applications} \label{applications}

The dynamic procedure developed for the evaluation of the model coefficient $c_\mu$ is based on the paradigm provided by the temporally filtered Navier-Stokes equations and on turbulence closures commonly employed in the modeled RANS equations. These are two key aspects for the developed coarse-grained approach. Indeed, from one side we have that the dynamic procedure explicitly pursues the temporal filtering formalism thus enabling a scale-resolving flow solution when the spatial and temporal resolution allow it. From the other side, we have that the baseline turbulence closure is a RANS model thus enabling, together with the large time-scale limit of the temporally filtered equations, a statistical flow solution when the spatial and temporal resolution is too coarse. In other words, the developed dynamic RANS approach represents an unifying closure with the potential of improving predictions of flow solutions ranging from the steady-state statistical to the unsteady scale-resolving ones, depending on the numerical settings adopted.

In light of the above arguments, the target of the developed dynamic RANS approach is to improve the practice in very-coarse simulations such as those commonly adopted in real-world CFD applications, namely steady-state statistical RANS and unsteady scale-resolving hybrid LES/RANS approaches. In particular, we foresee two main areas where the developed dynamic procedure is expected to provide a significant contribution. 1) Improvement of the accuracy of the solution with respect to classical RANS closures when adopting very coarse-grained and steady-state numerical settings. 2) Unlocking more accurate scale-resolving features than traditional hybrid LES/RANS approaches when the grid and spatial resolution allow them to be developed in coarse-grained unsteady numerical settings. As far as it concerns the latter, we foresee an improved ability in capturing laminar to turbulence transitional phenomena and in the responsiveness to changes in the free-stream turbulence boundary conditions. All these aspects will be quantitatively addressed through numerical simulations in the following sections.


\section{Transitional flow over a rounded leading-edge flat plate}
\label{sec_flowsetup}

The present section reports the first attempt to assess the properties of the proposed dynamic procedure. Different flow cases have been considered, such as the flow around a rectangular cylinder and the flow over a bump which are reported in appendix \S\ref{app_barc} and \S\ref{app_bump}, respectively. Here, we focus on the transitional flow over a rounded leading edge flat plate under different free-stream turbulence conditions, the so-called T3L test case series of the ERCOFTAC suite \citep{yang2001large, Crivellini20, cimarelli2020numerical}. The T3L case is chosen because it explicitly concerns flow phenomena of separation, transition and reattachment, which are intimately linked with the management of the near-wall physics, of the laminar to turbulence transition and of the physical dependence of such phenomena on the values of free-stream turbulence, the main prospects for the developed dynamic procedure. Simulations are performed by varying the grid resolution and the free-stream turbulence levels by keeping a fixed Reynolds number, $Re = U_0D/\nu =3333$ where $U_0$ is the free-stream velocity, $D$ the plate thickness and $\nu$ the kinematic viscosity.

\subsection{Numerical configuration}
\label{subsec:NC}

Simulations of the new modeling approach are performed by using the solver T-Flows (https://github.com/DelNov/T-Flows) \citep{hadvziabdic2022rational}. T-Flows is based on a finite volume method for collocated, unstructured and arbitrary grids. The transport scheme for momentum and turbulent quantities is SMART, a high-order upwind-biased scheme which maximizes accuracy within the limits required to preserve boundedness \citep{SMART}. Time integration is performed with the backward Euler scheme in first-order form for the stresses and in a second-order variant, used in iterative fashion, for the inertial term. The time step is adapted to obtain a condition $CFL \approx 1.5$. 

\begin{figure}
\centering
  \includegraphics[width=0.45\linewidth, trim=0 1.cm 0 0, clip]{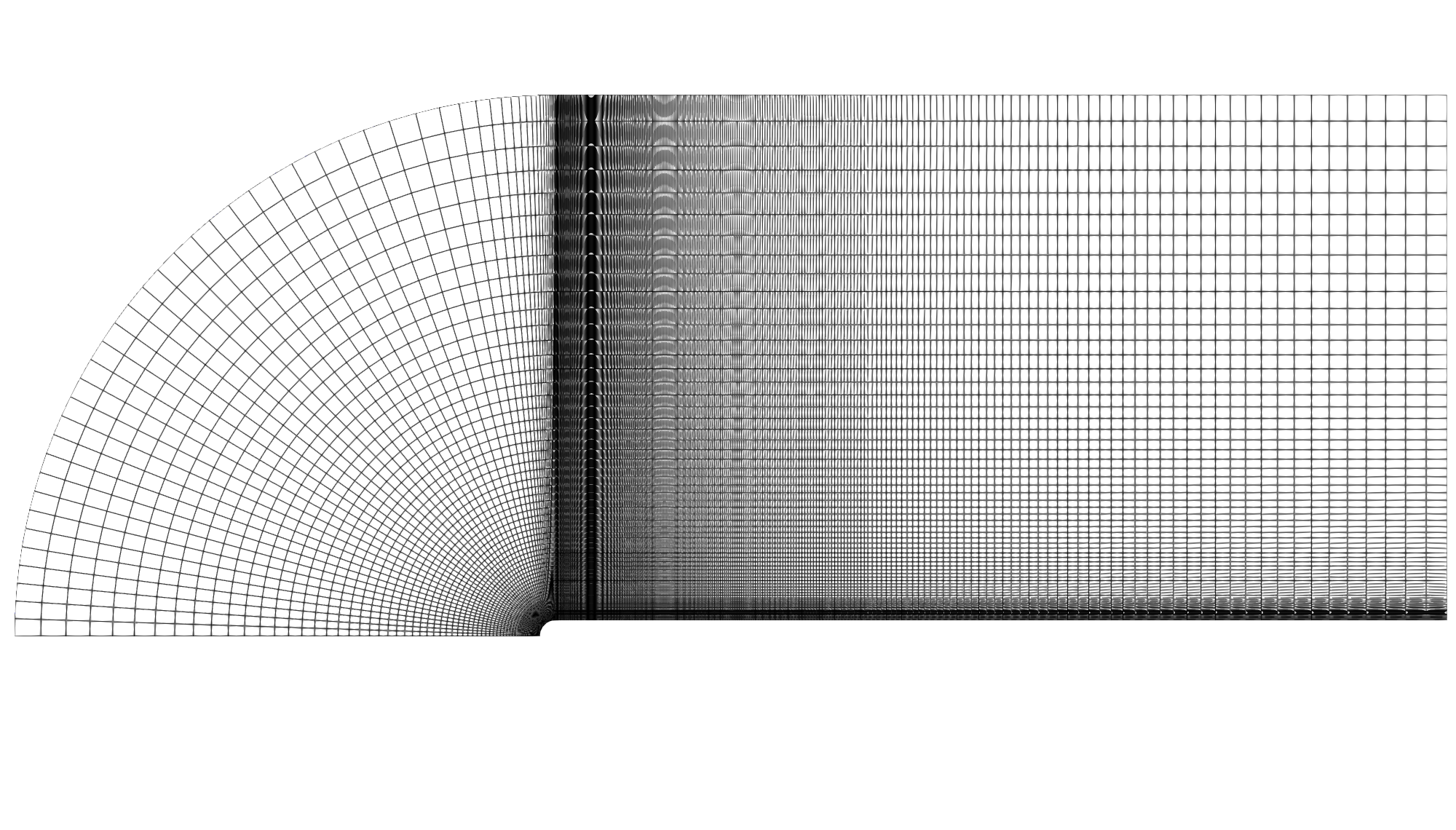}
  \includegraphics[width=0.45\linewidth, trim=0 1.cm 0 0, clip]{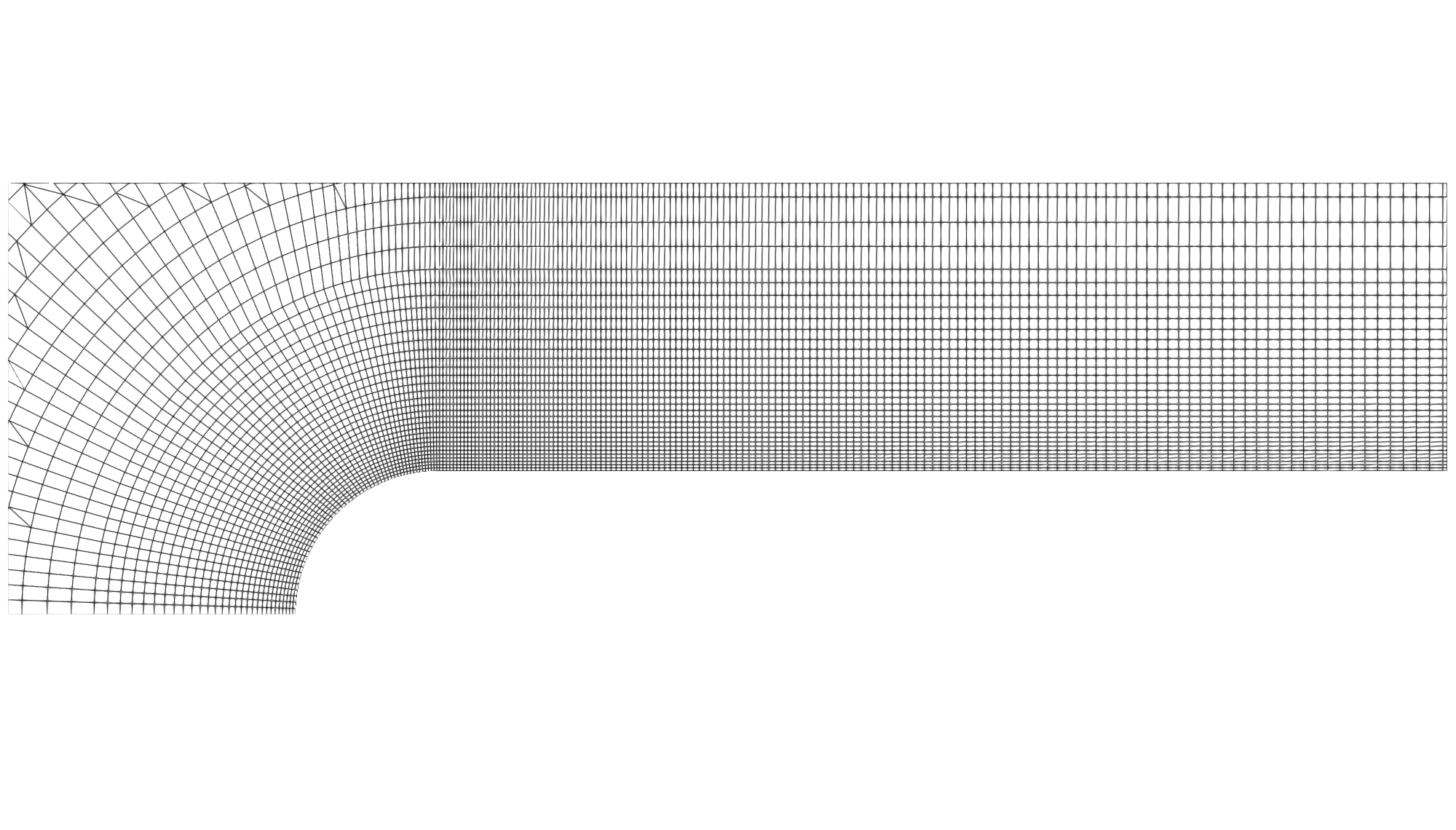}
  \includegraphics[width=0.45\linewidth, trim=0 1.cm 0 0, clip]{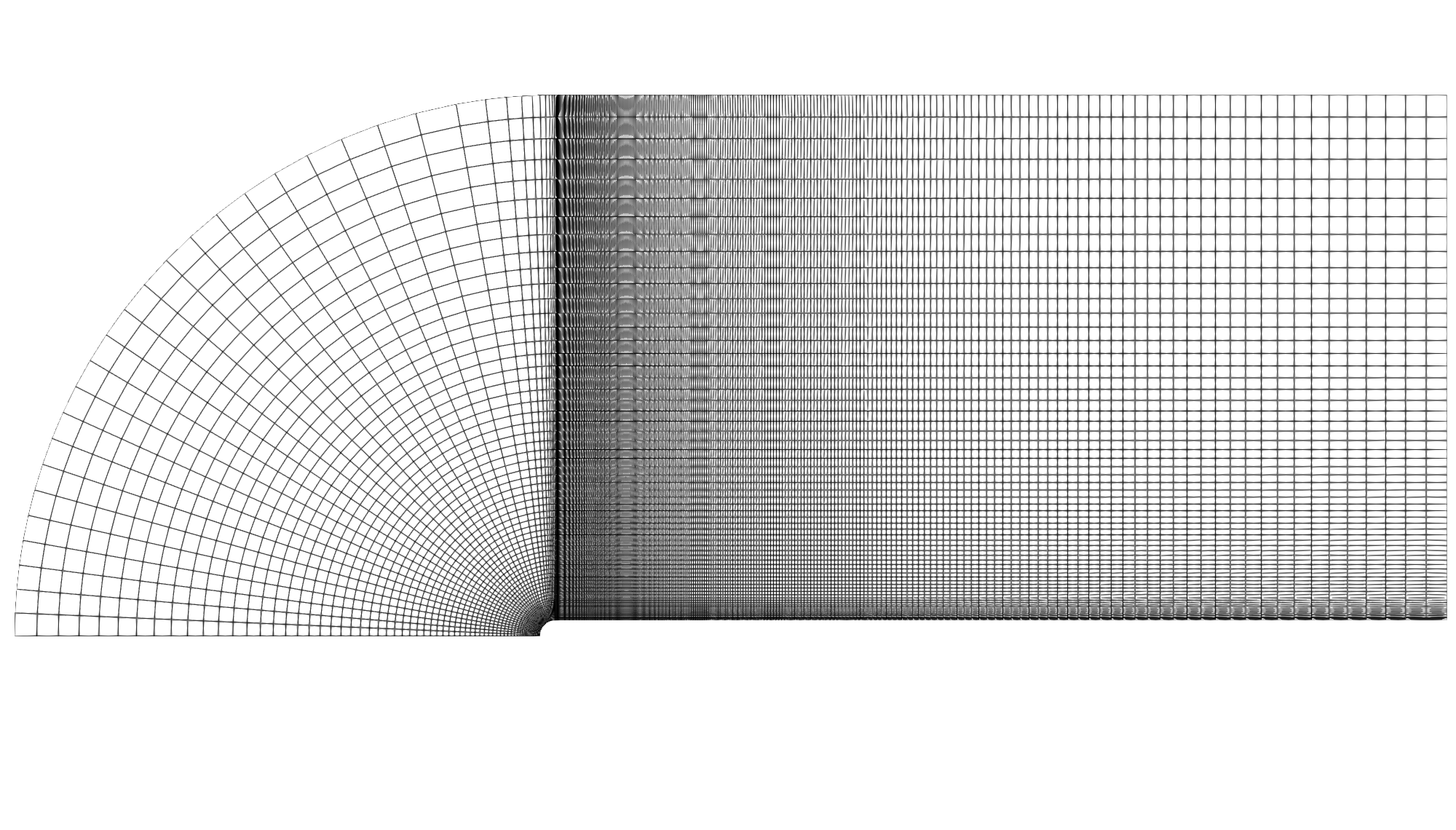}
  \includegraphics[width=0.45\linewidth, trim=0 1.cm 0 0, clip]{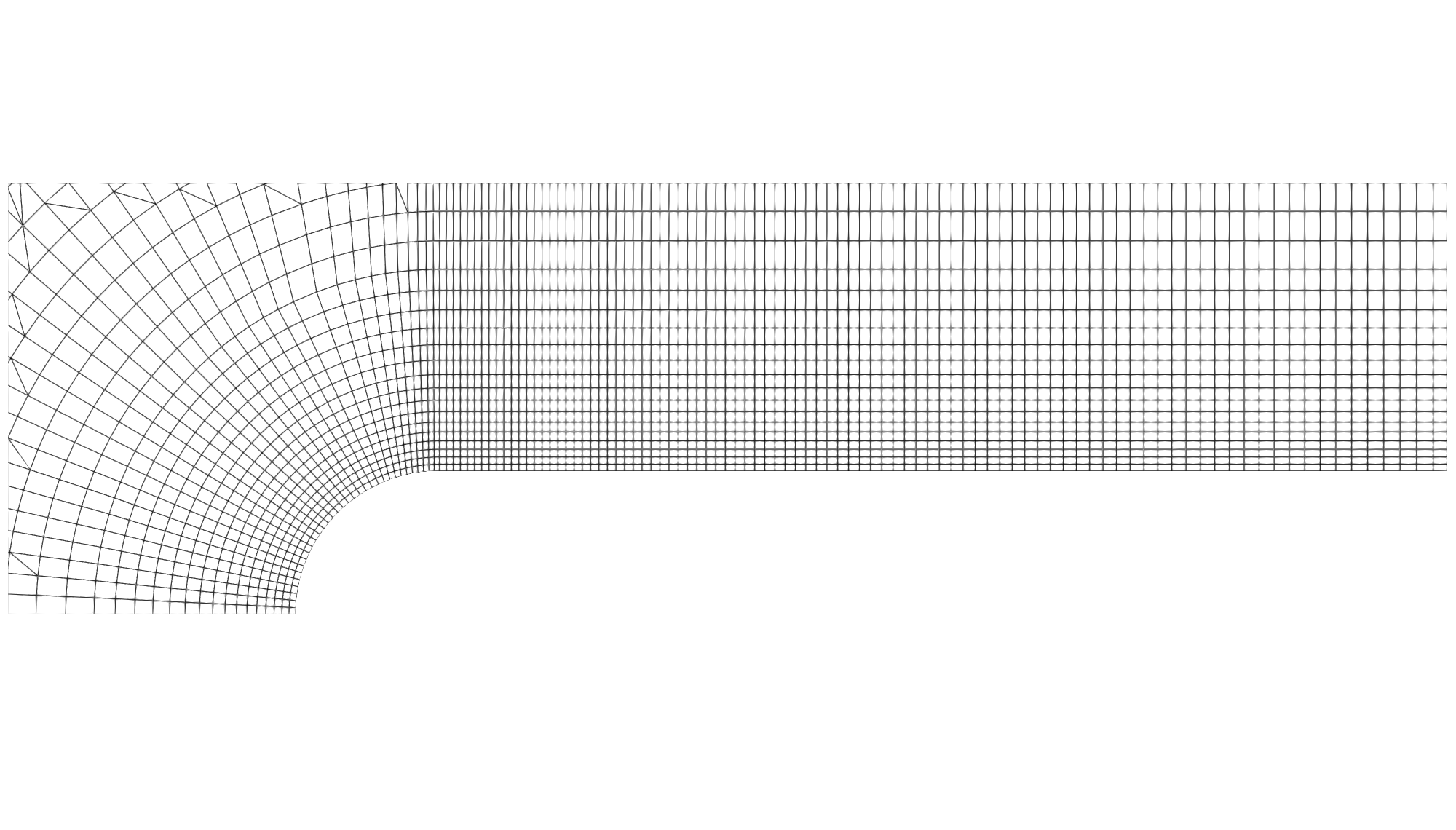}
  \caption{Finer (above) and coarser (below) grids for the T3L study. On the left, an overall view of the computational domain is reported while, on the right, a detailed view for $-1 \leq x/D \leq 4$ and $0 \leq y/D \leq 1.5$ is reported to better illustrate the grid resolution.}
  \label{fig:TMesh}
\end{figure}

By taking advantage of the statistical symmetries of the mean flow solution, only the top half of the flow domain is solved in line with common industrial approaches based on RANS. A fixed velocity is applied at the inlet while a free-slip condition is applied at the top and at the symmetry plane. A standard Neumann boundary condition is applied at the outlet in order to avoid the potential stability issues related to having a convective outlet adjacent to the finely resolved non-slip wall of the solid body. Finally, periodic boundary conditions are applied in the spanwise direction. The domain extends for $-17 \le x/D \le 28$, $0 \le y/D \le 17$ and $0 \le z/D \le 2$ where $(x,y,z)$ denote the streamwise, vertical and spanwise directions and the reference frame system is centered at the leading edge of the plate. Accordingly, $(u,v,w)$ will hereafter denote the streamwise, vertical and spanwise components of the velocity vector.

The geometric characteristics of the rounded leading edge flat plate offer the possibility of having a structured layout for the grid by adopting a C-type inlet. Two different grids are considered in order to address the sensitivity of the flow solution to spatial discretization, see figure \ref{fig:TMesh}. Both grids have a first layer of cells around the plate whose thickness is, on average, smaller than one friction length, thus supporting the use of a low-Reynolds formulation for the baseline $k-\varepsilon$ closure. The T3L case exhibits a laminar separation around the end of the circular leading edge and a reattachment at streamwise positions varying with the turbulent intensity of the free stream. Accordingly, both grids exhibit a stretching leading to finer spatial resolutions by moving towards the plate in the vertical direction and towards the leading edge in the streamwise direction. The total number of computational elements in the finer grid is 1.03M while for the coarser grid spatial resolution is roughly halved in all 3 directions, resulting in 0.34M numerical volumes. Overall, the spatial resolution adopted for the finer grid evaluated in a refinement box extending for $0.5 \le x/D \le 1.2$ and $0.5 \le y/D \le 9.3$, is $\Delta x/D = 0.012 \div 0.1$, $\Delta y/D = 0.01 \div 0.046$ and $\Delta z/D = 0.05$ while for the coarser grid is $\Delta x/D = 0.023 \div 0.11$, $\Delta y/D = 0.023 \div 0.073$ and $\Delta z/D = 0.1$. In the present work, we also perform 2D steady-state simulations. For these simulations, the grid employed has the same $x-y$ distribution of the finer 3D grid and counts 25700 computational elements.

\subsection{Dynamic procedure settings}
\label{sec_dyn_settings}

\subsubsection{Baseline closure}
\label{sec_closure_setting}

The dynamic RANS procedure developed in section \S\ref{dynev} represents a new class of turbulence closures, as the procedure can be adapted to any type of two-equation turbulent viscosity closures. In this first attempt, we assess its performance by using the standard low-Re $k-\varepsilon$ model by \citet{abe1995new} as underlying closure, in line with the theoretical development reported in section \S\ref{dynev}. A complete assessment of its performance including the effect of a change of the baseline closure is left to future more technical works.

It is important to highlight that standard implementations of the $k-\varepsilon$ model make use of a damping function $f_\mu$ to limit the value and improve the scaling of eddy viscosity in the near-wall region, i.e. $\nu_\tau = f_\mu c_\mu k_\tau^2 /\epsilon_\tau$. This correction to the standard $k-\varepsilon$ approach is no longer necessary when adopting the developed dynamic procedure. As reported in appendix \S\ref{app_effects}, we performed tests with and without damping function $f_\mu$ and we found that the dynamic procedure automatically adjusts the eddy viscosity values to the different flow regions thus producing more physically accurate results without the use of artificial corrections to the eddy viscosity field. Accordingly, all the simulations reported in the present work are performed by using the standard low-Re $k-\varepsilon$ model as baseline closure without damping functions for the eddy viscosity. Note that the damping function $f_\mu$ must not be confused with the damping function $f_\epsilon$ that adjusts the near-wall scaling of the destruction term of the $\varepsilon$ equation (\ref{eq_eps}), which is instead still required if the equations are to be integrated down to the wall, as they are in the present simulations.

\subsubsection{Filter kernels and widths}

Apart from the baseline closure, the new dynamic RANS approach presents another important degree of freedom that can lead to different type of implementations: the kernel of the second test filter and its width $T$. Given the dual nature of the dynamic RANS procedure that ranges from statistical to scale-resolving approaches, the selection of the filter kernel is not straightforward and is addressed here for both approaches separately.

When seeking for a scale-resolving unsteady solution, the simplest choice for the test filter kernel is that given by a backward moving average,
\begin{equation}
\langle \beta \rangle_{T} (N, t) = \frac{1}{N\Delta t} \int_{t-N\Delta t}^t \!\! \beta (t') \, dt'
\label{eq_moving_ave}
\end{equation}
where $N$ is the number of time steps used for computing the moving average and, hence, $T = N \Delta t$. As shown in appendix \S\ref{app_effects}, tests have been performed to address the sensitivity of the flow solution to the width of the test filter. We found that the solution provided by the new dynamic RANS approach is largely insensitive to values of the test filter width ranging from $N=2$ to $N=5$. Hence, in the present work, we report results of the dynamic RANS approach only for $N=2$, i.e. for the simplest and computationally cheapest implementation of the dynamic procedure.

When seeking a statistical solution, the definition of the test filter operator is less straightforward. Indeed, the common CFD practice for computing a statistical solution is to use steady-state solvers in order to save computational costs. Steady-state solvers commonly adopt iterative methods that are seeking for a convergence of the flow solution. This convergence is usually achieved by RANS closures due to the high intensity of the modeled stresses introduced into the solution that allows to rapidly reach a statistical solution provided that the separation of scales of the problem is not too large. On the contrary, when the problem under consideration is characterized by a large separation of scales, the intensity of the modeled stresses of RANS closures is not high enough to reach a statistical solution and the iterative solutions of the solver are found to fluctuate. This last scenario is what characterize the dynamic RANS solution when computed using steady-state solvers. Indeed, we have that the modulation of the modeled stresses given by the dynamic procedure leads to variable iterative solutions also in flow problems characterized by a small separation of scales. Because in steady-state solvers the iterative solutions are in general not correlated being not separated by a physical time, a possible choice for the test filter kernel is to use a cumulative algebraic average of the iterative solutions. For a generic quantity $\beta$, the resulting test filter operator reads
\begin{equation}
\langle \beta \rangle_{T} (it_0, it_n) = \frac{1}{it_n-it_0} \sum_{it_0}^{it_n} \beta (it)
\label{eq_accum_ave}
\end{equation}
where $it_0$ is the first iteration used for the algebraic average accumulation and $it_n$ is the last one. In the present work, we start the dynamic RANS steady-state simulations from an initial iterative solution $it_0$ obtained by running the baseline closure without dynamic procedure.

\subsubsection{Clipping procedure}

It is worth noting that the dynamic procedure (\ref{eq_dyn_eq}) can yield negative values for the model coefficient $c_\mu$ and, hence, of eddy viscosity with obvious repercussions for the numerical stability of the solution. For this reason, we adopt a clipping procedure that limits the dynamic coefficient's value to the range $0 \leq c_\mu \leq 0.2$. Extensive tests have again been performed in order to assess the sensitivity of the solution to the imposed upper bound value of $c_\mu$. As shown in appendix \S\ref{app_effects}, a small sensitivity of the solution to this value has again been found thus further supporting the robustness of the developed dynamic procedure.

\subsection{Free-stream turbulence}
\label{sec_fs_bc}

Transition, separation and reattachment are key physical features of the T3L case and many other applications and are known to significantly depend on the free-stream turbulence conditions. The common practice in real-world CFD applications is to prescribe the free-stream turbulence conditions through the inlet boundary conditions of the modeled turbulent kinetic energy and dissipation, $k_{in}$ and $\varepsilon_{in}$. Since the target of the present work is to improve the CFD practice, we adopt the same approach here. However, this choice makes the prescription of the free-stream turbulence conditions rather complex. Indeed, the transported quantities experience a decay before impinging to the body that, without loss of generality, can be for example modeled as
\begin{equation}
  \left .
    \begin{split}
      k(x)=k_{in} \left[ 1+\Big (C_{\varepsilon2}-1 \Big) \frac{\varepsilon_{in}}{k_{in}} \frac{x}{U_{0}} \right]^{-\frac{1}{C_{\varepsilon2}-1}}\\
      \varepsilon(x)=\frac{k(x)\varepsilon_{in}}{k_{in}\left( 1+\frac{\varepsilon_{in}}{k_{in}}\frac{x}{U_{0}} \right)}
    \end{split}
  \right \}
\end{equation}
However, the actual decay changes depending on the numerical resolution, numerical schemes and turbulence closure. Hence, a fine tuning of the boundary conditions is required to prescribe a specific free-stream condition impinging to the body. To note that the free-stream turbulence conditions are actually unknown in most applications, since they depend on atmospheric conditions, wakes of other objects, and other complex phenomena. Overall, the difficulties in replicating the exact same free-stream turbulence in different simulations, together with the limited relevance of the actual conditions (as they are generally problem variables), render the study of precise free-stream turbulence conditions and the associated fine-tuning of the inlet boundary conditions irrelevant for real-world applications. The actual value for applications is for the simulations to correctly capture the sensitivity of the flow solution to variations of the free-stream turbulence, in order to address the robustness of the aerodynamic design to the ambient conditions. In the present work, we stick to the real-world practice by imposing the same combinations of $k_{in}$ and $\varepsilon_{in}$ for all closures, rather than iteratively adjusting them individually to replicate the free-stream conditions that are eventually perceived by the rounded leading edge flat plat. Accordingly, the performance of the different modelling approaches will not be evaluated in terms of the precision in replicating the reference dataset, but rather of their ability to capture the physics of the different flow regions and their dependence on changes of the free-stream turbulence conditions. The latter are here evaluated by computing the free-stream turbulence intensity and the free-stream turbulence integral length in a plane $x=-D$,
\begin{equation}
    \begin{split}
      &I=\frac{\sqrt{2 r \, k / 3}}{U_0}\\
      &\lambda=\frac{\left (r \, k \right )^{3/2}}{\varepsilon}
    \end{split}
\end{equation} 
Here, $r = (c_\mu / c_\mu^*)^{1/2}$ is a scaling factor deemed essential to take into account the fact that in the dynamic RANS procedure here developed, the actual free-stream turbulence acting in the momentum equations is modulated by the value of the model coefficient $c_\mu$, contrary to more classical approaches where the model coefficient $c_\mu^*$ is prescribed as a constant.


\section{Scale-resolving approach: 3D unsteady simulations}
\label{sec_3D_results}

We start the assessment of the developed dynamic RANS procedure by considering its performance when scale-resolving simulation settings are adopted. The specific settings of the dynamic procedure have been already reported in section \S\ref{sec_dyn_settings} but we remind here that for scale-resolving simulations we adopt the backward moving average (\ref{eq_moving_ave}) as test filter operator. To have a direct indication of the effect of the dynamic procedure on the solution, results will be compared with those obtained with the baseline model, the standard $k-\varepsilon$ approach with $c_\mu^* = 0.09$. To address the scale-resolving capabilities of the dynamic procedure, results will also be compared with a hybrid LES/RANS approach \citep{frohlich2008hybrid,spalart2021hybrid}. The hybrid closure used in the present work employs the Smagorinsky model in the free-flow regions and relies on the $k$-$\varepsilon$-$\zeta$-$f$ model in the near-wall regions \citep{hanjalic2004robust}. For further details about the hybrid LES/RANS model, the reader is referred to \citet{delibra2010vortex}. Finally, data from high-fidelity implicit-LES simulations performed by \citet{Crivellini20} will be considered as a reference.

\begin{figure*}
\centering
  \includegraphics[width=0.45\linewidth]{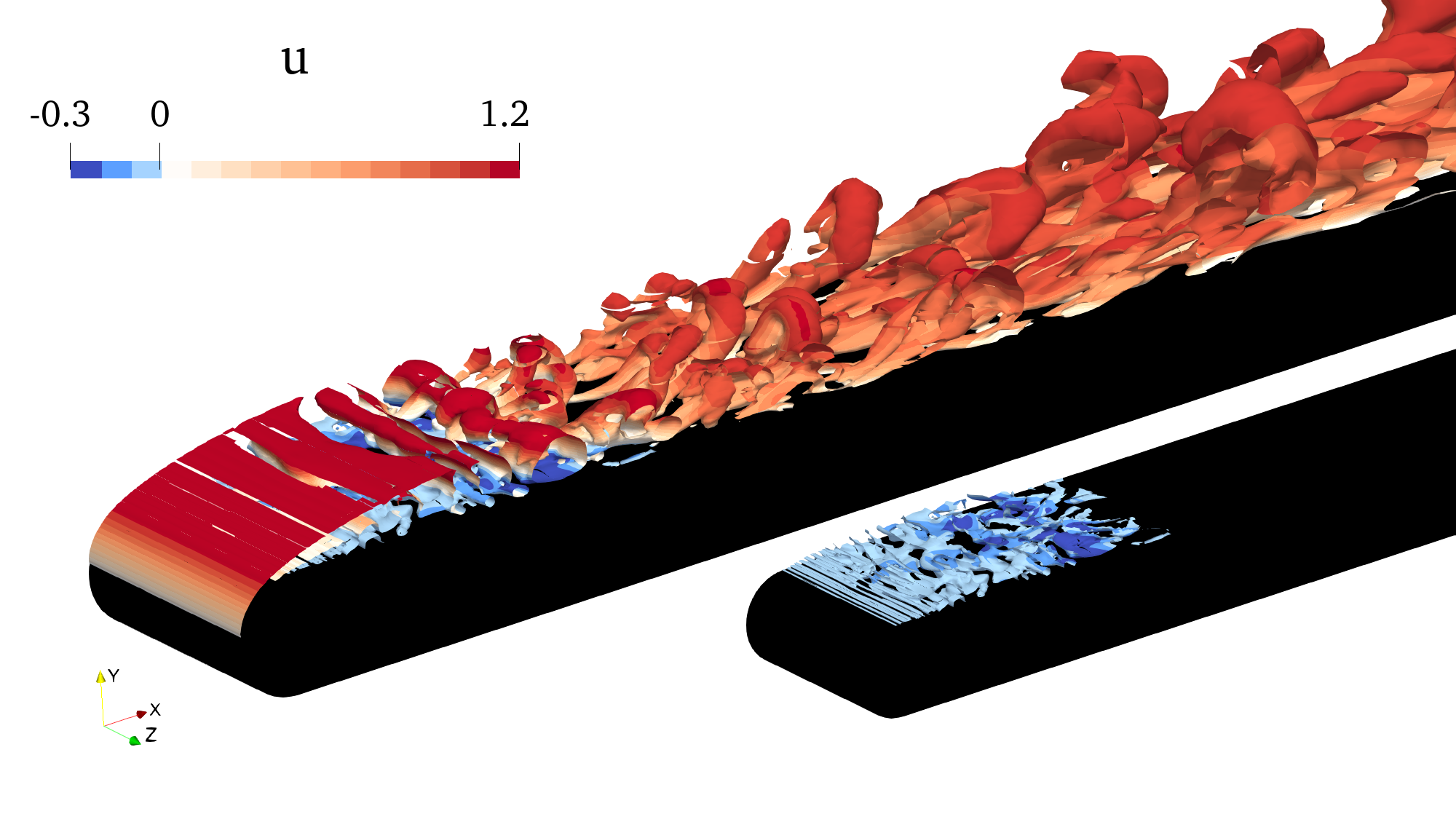}\mylab{-80mm}{30mm}{(\aaa)}  \hspace{0.2cm}\includegraphics[width=0.45\linewidth]{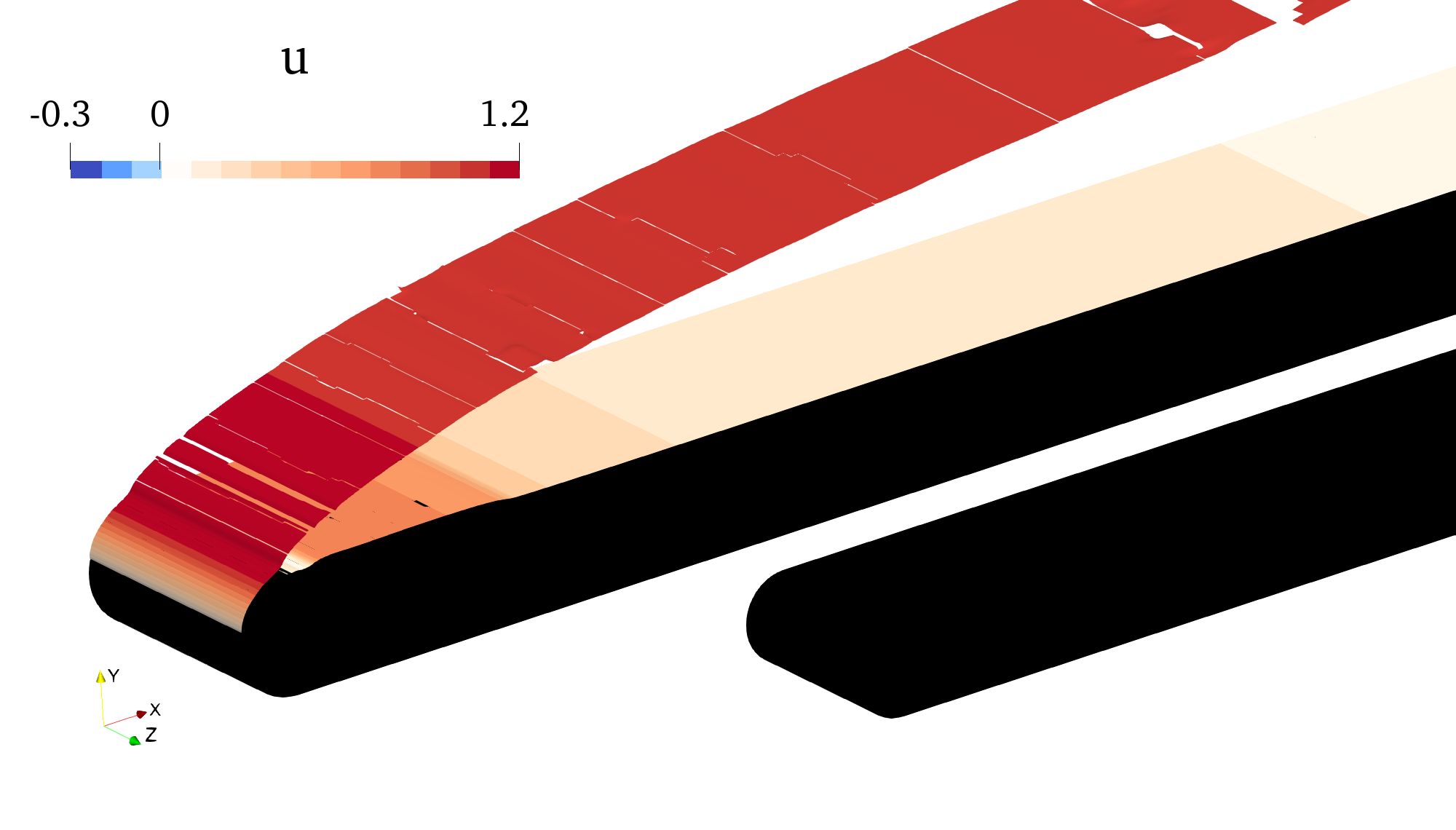}\mylab{-80mm}{30mm}{(\bbb)}
\vspace{0.4cm}\includegraphics[width=0.45\linewidth]{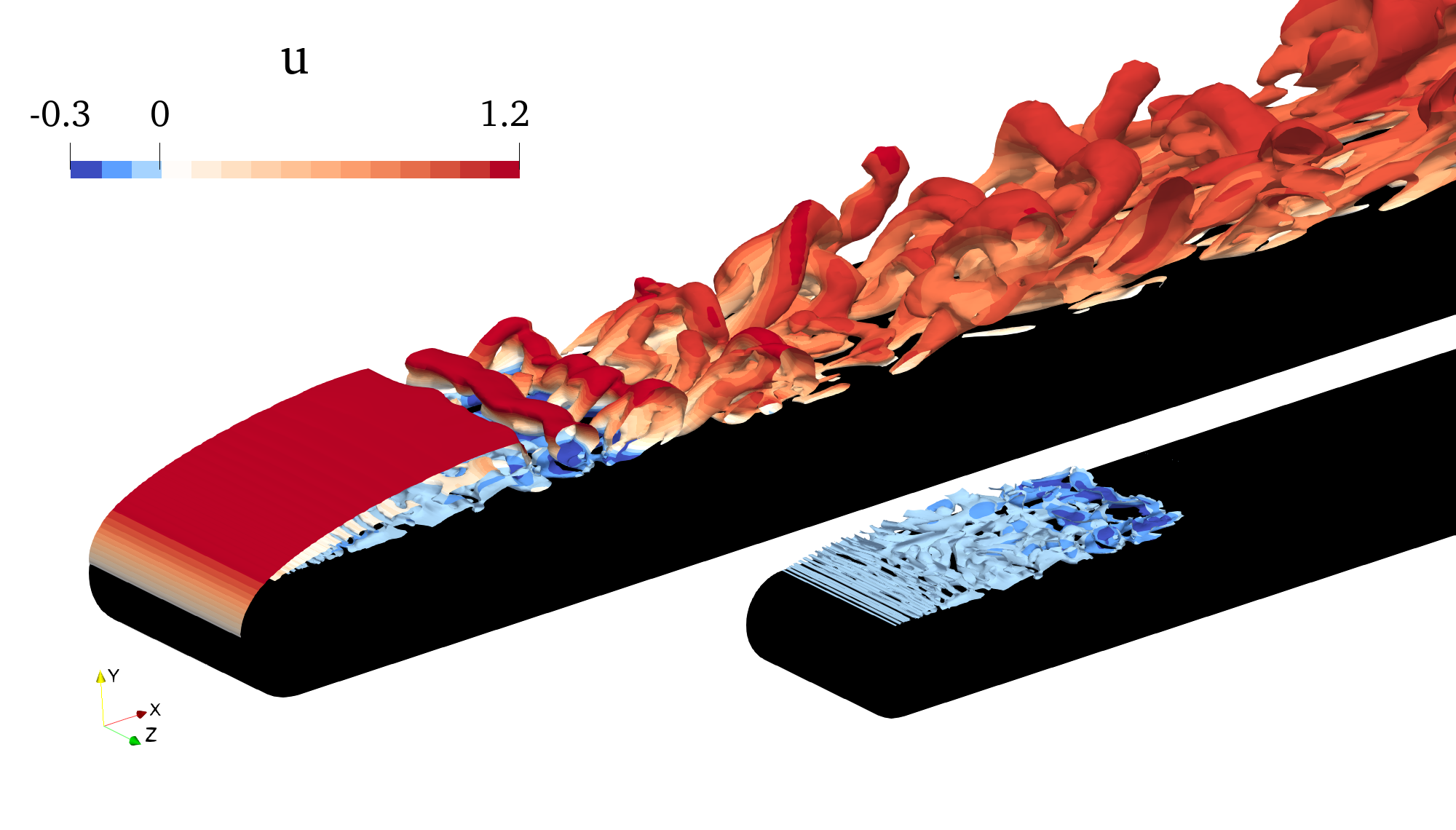}\mylab{-80mm}{30mm}{(\ccc)}
  \caption{Fine-grid scale-resolving simulations of the transitional flow over a rounded leading-edge flat plate. Iso-surfaces of the Q-criterion \citep{hunt1988eddies}, $Q = 0.15$, colored with the streamwise velocity for (a) the new dynamic RANS closure with $I = 0.35\%$, (b) the standard $k-\varepsilon$ model with $I = 2.3\%$ and (c) the hybrid LES/RANS model with $I = 0\%$. In the inset plots, iso-surfaces with positive streamwise velocity are hidden to reveal the structures of the reverse boundary layer within the main recirculating bubble.}
  \label{fig_inst_fields}
\end{figure*}

Different free-stream turbulence conditions are simulated by varying the free-stream boundary conditions of the modeled turbulent kinetic energy and dissipation, $k_{in}$ and $\varepsilon_{in}$ respectively. By following the procedure reported in section \S\ref{sec_fs_bc}, the imposed values of $k_{in}$ and $\varepsilon_{in}$ are designed in order to systematically change the turbulence intensity $I$ perceived by the body while keeping the integral length almost constant $\lambda = D/2$. In particular, three different combinations of the boundary conditions will be considered, yielding three flow cases, at low, medium and high free-stream turbulence intensity $I$. Statistics are computed by performing a spatial average in the spanwise direction and a temporal average based on 60 flow samples gathered every $6L/U_0$ times after reaching a fully developed flow solution. The resulting average operator is hereafter denoted as $\langle \cdot \rangle$.

\subsection{Instantaneous flow pattern}

The instantaneous flow pattern taken by the new dynamic RANS approach is reported in figure \ref{fig_inst_fields}(a). A fully unsteady and 3D flow pattern is reproduced. First, an almost laminar shear layer is developed that takes origin from the separation of the boundary layer around the beginning of the flat portion of the plate. The shear layer is then found to develop Kelvin-Helmholtz instabilities taking the form of spanwise vortex tubes spanning the entire domain width. By moving downstream, the spanwise vortices undergo a further instability in the form of a lift up and stretching thus forming hairpin-like structures arranged in a staggered manner \citep{cimarelli2018structure}. After reattachment, these hairpin-like structures are further stretched by mean shear thus developing streamwise vortices which are known to induce high- and low-speed streaks that dominate the downstream development of the attached boundary layer. To note that within the recirculating bubble, a reverse boundary layer also takes place that as highlighted in the inset of figure \ref{fig_inst_fields}(a) is also fully turbulent.

As shown in \ref{fig_inst_fields}(b), the baseline $k-\varepsilon$ model completely misses all the degrees of freedom of the solution. Indeed, a stationary 2D solution is reproduced that lacks the turbulent structure that would otherwise be supported by the grid resolution, as demonstrated by the dynamic RANS closure. Such a scale-resolving feature of the dynamic RANS procedure is indeed recovered when considering scale-resolving methods such as the hybrid LES/RANS approach reported in \ref{fig_inst_fields}(c). In conclusion, the spatial resolution adopted supports a large number of degrees of freedom in the solution that are unlocked by the new dynamic RANS approach and by the hybrid LES/RANS model, and which would otherwise be missed by using standard statistical closures such as the $k-\varepsilon$ model. 

\begin{figure*}
\centering
  \includegraphics[width=0.45\linewidth]{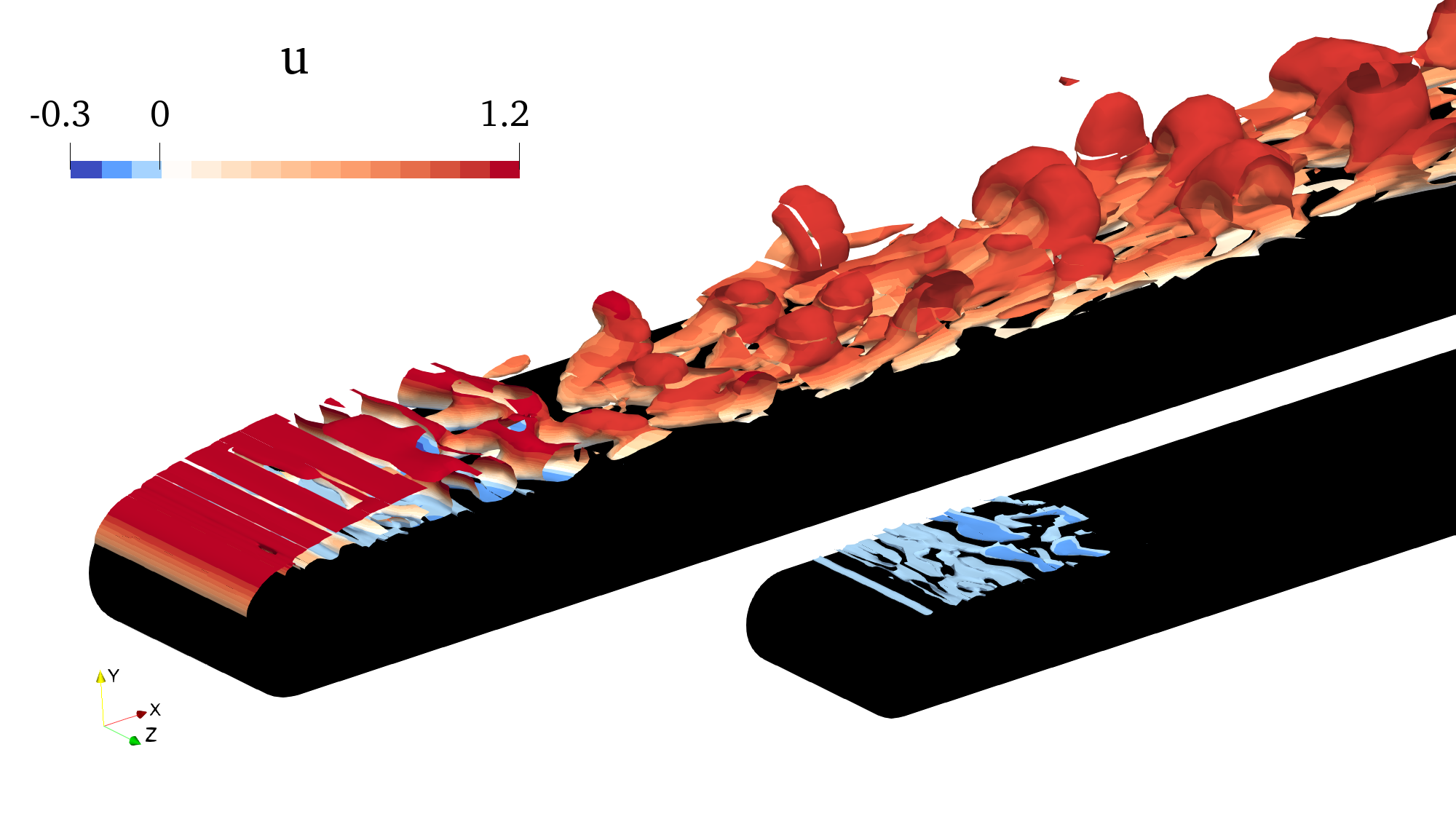}\mylab{-80mm}{30mm}{(\aaa)}  \hspace{0.2cm}\includegraphics[width=0.45\linewidth]{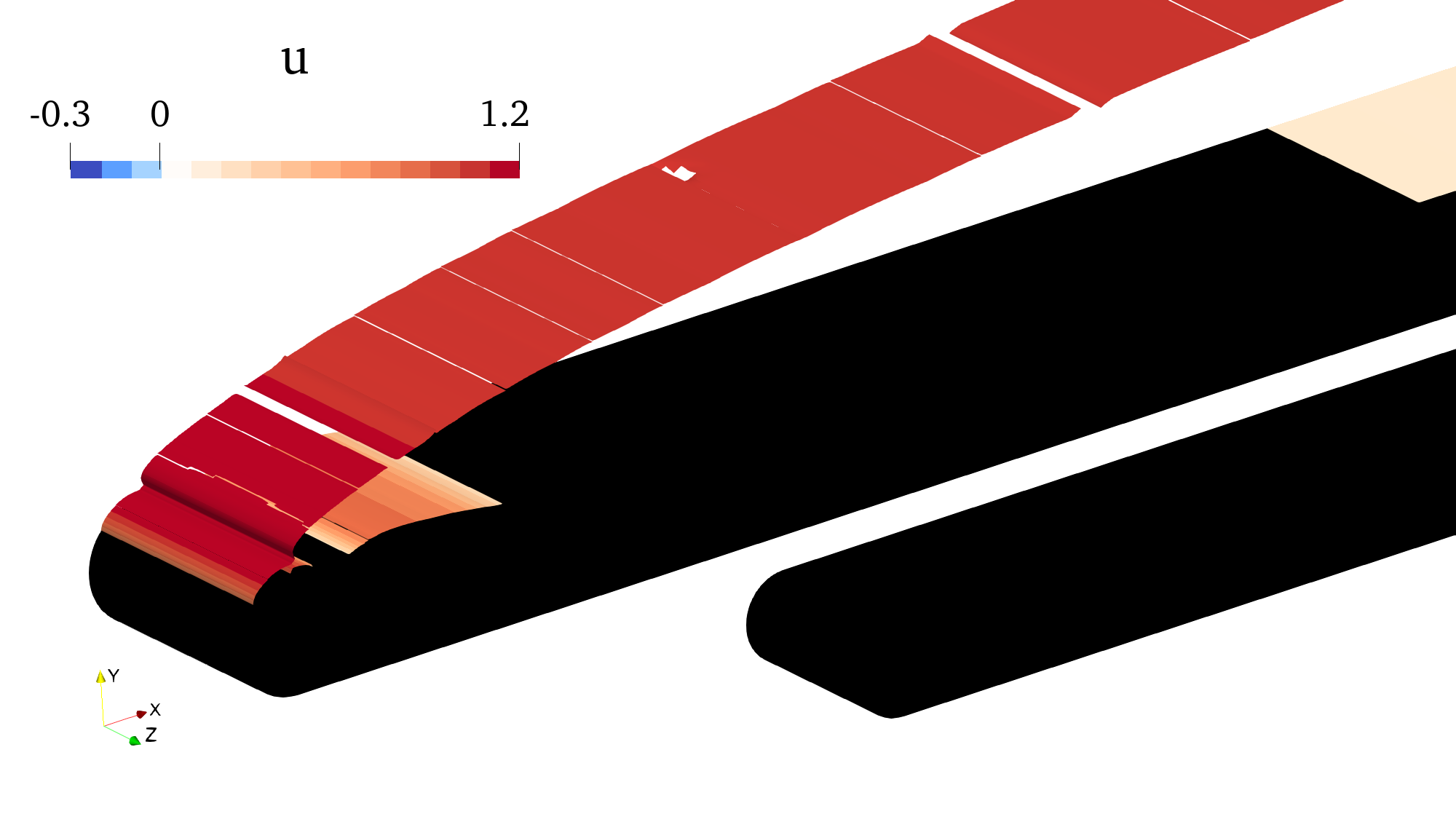}\mylab{-80mm}{30mm}{(\bbb)}
\vspace{0.4cm}\includegraphics[width=0.45\linewidth]{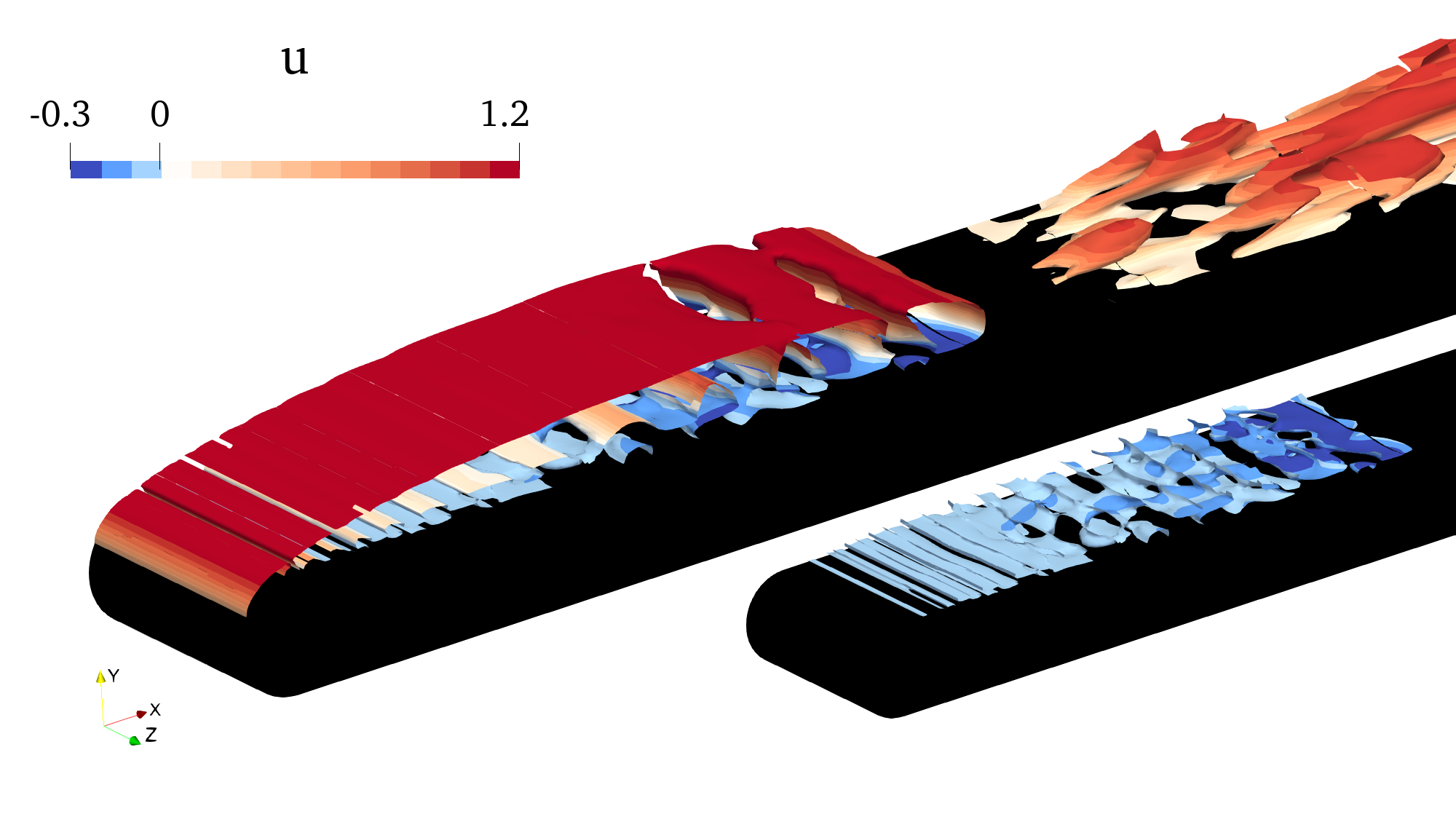}\mylab{-80mm}{30mm}{(\ccc)}
  \caption{Coarse-grid scale-resolving simulations of the transitional flow over a rounded leading-edge flat plate. Iso-surfaces of the Q-criterion \citep{hunt1988eddies}, $Q = 0.15$, colored with the streamwise velocity for (a) the new dynamic RANS closure with $I = 0.5\%$, (b) the standard $k-\varepsilon$ model with $I = 2.5\%$ and (c) the hybrid LES/RANS model with $I = 0\%$. In the inset plots, iso-surfaces with positive streamwise velocity are hidden to reveal the structures of the reverse boundary layer within the main recirculating bubble.}
  \label{fig_inst_fields_coarse}
\end{figure*}

As shown in figure \ref{fig_inst_fields_coarse}(a), by coarsening the spatial resolution, the flow solution of the dynamic RANS model still reproduces the relevant flow structures highlighted so far for the finer grid. The only difference is simply given by the fact that the smaller flow structures  reproduced by the finer grid are not anymore captured for obvious reasons of spatial resolution. Such a robustness of the dynamic RANS approach to the coarsening of the resolution is not observed for the hybrid LES/RANS approach. Indeed, as shown in figure \ref{fig_inst_fields_coarse}(c), the flow solution of the hybrid LES/RANS method suffers a strong deterioration of the resolved flow physics that is not limited to the sole absence of fine scale structures in the flow solution. It consists also in a strong degradation of relevant flow features such as the large-scale dynamics of the recirculating bubble whose length scale is widely overestimated and whose structure is reduced to almost two-dimensional spanwise rolls.

The robustness of the dynamic RANS closure to the coarsening of the spatial resolution is a remarkable feature rendering it particularly interesting for CFD applications in real-world problems. This relevant property stems from the fact that the developed model is a RANS closure in nature wherein the dynamic procedure locally adapts to the spatial resolution by enabling scale-resolving features also in ambiguous grids. Indeed, when coarsening the grid resolution, larger portions of the flow domain call for an increase of the modeled stresses that would asymptotically converge to that provided by statistical approaches typical of pure RANS closures. These transitional and fully modeled regions are better captured by the developed dynamic RANS closure thanks to its ability in improving the flow solution also in statistical approaches as it will be demonstrated in section \S\ref{sec_2D_results} through the analysis of 2D steady simulations.

\subsection{Dynamic model coefficient}

\begin{figure*}
  \centering
  \includegraphics[width=0.9\linewidth]{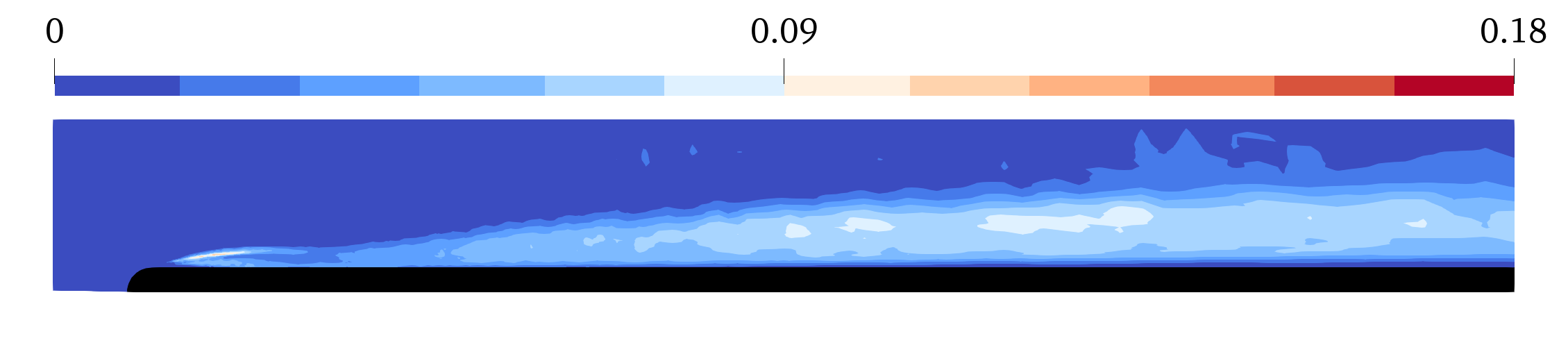}
  \caption{Average behaviour of the model coefficient $\langle c_\mu \rangle$ obtained by the dynamic RANS closure for the scale-resolving 3D unsteady simulations performed in the fine grid. Note that shades of blue denote areas where the model coefficient is on average lower than the classical value $c_\mu = 0.09$, whereas shades of red correspond to the few instances where it is higher.}
\label{fig_coeff}
\end{figure*}

In order to address the ability of the dynamic RANS closure to automatically adapt to different flow physics and in enabling scale-resolving features, we analyse here the behaviour of the model coefficient $c_\mu$ shown in figure \ref{fig_coeff}. In the leading-edge shear layer, the dynamic procedure reproduces its transitional physics. It consists of small values of the model coefficient in the first laminar part and in a gradual downstream increase in its intensity --up to a peak of $c_\mu = 0.12$-- that matches the development of increasingly higher levels of turbulence. As a result the dynamic RANS closure is able to reproduce laminar to turbulence transition. This property of the dynamic RANS closure is of overwhelming relevance for actual CFD applications. As shown in figure \ref{fig_coeff}, the model coefficient is found to be activated by the dynamic procedure also in the reverse boundary layer within the leading-edge recirculating bubble and in the downstream development of the attached boundary layer. Also in this case, the dynamic procedure is found to reproduce the physics of the problem that for boundary layers consists in a strongly inhomogeneous phenomenology of momentum transport that is dominated by viscous diffusion in the near-wall region and by turbulence mixing in the bulk of the flow. In fact, the model coefficient $c_\mu$ is found to dynamically adapt to these different physics of the boundary layer. In particular, the model coefficient is found to assume values similar to the ones classical employed in standard $k-\varepsilon$ implementations, $c_\mu = 0.09$, in the bulk of the flow where turbulence mixing dominates momentum transport. From these values, the model coefficient is found to decrease by moving towards the wall thus naturally adapting to the low Reynolds number behaviour of the near-wall region. This property is again of relevance for actual CFD applications. Indeed, as already stated in section \S\ref{sec_closure_setting}, this scaling of $c_\mu$ allows to avoid the use of near-wall damping functions whose use is critical especially in separating and reattaching flows typical of the applications. In conclusion, the dynamic RANS closure is able to adapt to the strongly inhomogeneous physics of the different flow regions by activating the dynamic coefficient $c_\mu$ in the fully turbulent regions and by suppressing it in the laminar viscosity-dominated ones even in flow regions characterized by high shear rates such as the very near-wall region and the initial laminar stage of the leading-edge shear layer.

\subsection{Friction coefficient}

\begin{figure}
  \centering
  \includegraphics[width=0.95\linewidth, trim=0 1.5cm 0 0, clip]{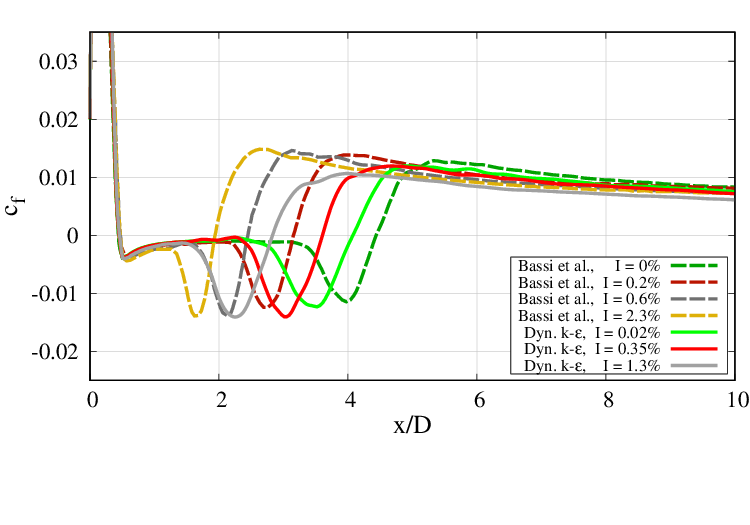}\mylab{0mm}{40mm}{(\aaa)}
  \includegraphics[width=0.95\linewidth, trim=0 1.5cm 0 0, clip]{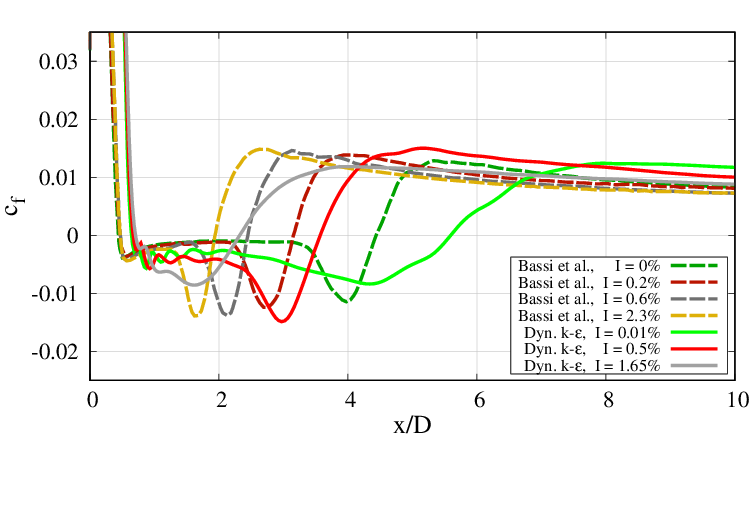}\mylab{0mm}{40mm}{(\bbb)}
  \caption{Average friction coefficient for the scale-resolving 3D unsteady simulations performed in (a) the fine and (b) coarse grids of the new dynamic RANS closure compared to the high-fidelity implicit-LES simulations performed by \citet{Crivellini20}.}
\label{fig_friction_1}
\end{figure}

The ability to deal with laminar flows and predict transition is one of the main prospects for the developed dynamic RANS closure. In this respect, the laminar separation, transition and turbulent reattachment events that characterize the T3L flow case make it very suitable to test the ability of closures in addressing these phenomena. In order to quantitatively assess the capability of the different turbulence closures in accurately capturing such flow features we report the behaviour of the friction coefficient $c_f$. As shown in figure \ref{fig_friction_1}(a), the new dynamic RANS approach is found to reproduce the phenomena of laminar separation, transition and turbulent reattachment and to capture their dependence on the value of free-stream turbulence. In particular, the shape of the friction coefficient profile reproduces that exhibited by the high-fidelity implicit-LES simulations performed by \citet{Crivellini20} thus showing that the physics of the phenomena involved in the flow system is nicely captured. Furthermore, the expected shrinking of the separation bubble and reduction of the reattachment length for increasing free-stream turbulence levels is also nicely reproduced thus showing that the dynamic RANS closure allows to capture flow phenomena such as transition that are sensitive to the free-stream turbulence level. When coarsening the resolution, the resolved physics deteriorates as shown in figure \ref{fig_friction_1}(b). However, the main features of the flow are still captured, namely the gross shape of the friction coefficient profile and its dependence on the free-stream turbulence.

\begin{figure}
  \centering
  \includegraphics[width=0.95\linewidth, trim=0 1.5cm 0 0, clip]{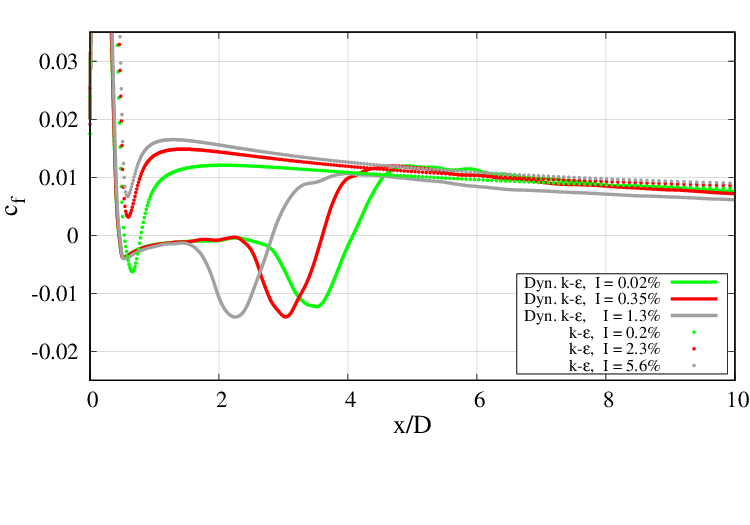}\mylab{0mm}{40mm}{(\aaa)}
  \includegraphics[width=0.95\linewidth, trim=0 1.5cm 0 0, clip]{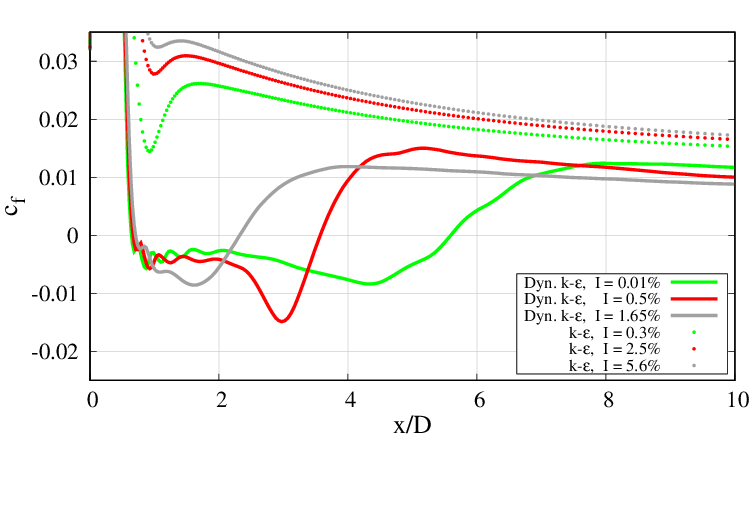}\mylab{0mm}{40mm}{(\bbb)}
  \caption{Average friction coefficient for the scale-resolving 3D unsteady simulations performed in (a) the fine and (b) coarse grids of the new dynamic RANS closure compared to the standard $k-\varepsilon$ approach.}
\label{fig_friction_2}
\end{figure}

The degree of improvement of the physics captured by the new modeling approach becomes apparent when comparing the behaviour of the friction coefficient with that obtained by using the baseline $k-\varepsilon$ approach. As shown in figure \ref{fig_friction_2}(a), the standard $k-\varepsilon$ model completely fails in capturing the laminar separation. Actually, the boundary layer developing at the leading edge is always turbulent and, as a consequence, the flow does not separate with except to a very small recirculating bubble for the flow case at the lower free-stream turbulence level. The scenario becomes even worse when coarsening the resolution. As shown in figure \ref{fig_friction_2}(b), the physics of the leading-edge boundary layer is so poor that affects the entire downstream development of the flow. Overall, the poor performances of the baseline $k-\varepsilon$ model are related to its well-known tendency to over-predict eddy viscosity in areas of the flow which should be laminar and transitional. As a consequence, the separation of the leading-edge boundary layer is almost always prevented, rendering the $k-\varepsilon$ model unsuitable for transitional separating and reattaching flows. In this context, the adoption of the present dynamic procedure to the $k-\varepsilon$ model makes it possible to solve all these problems.

\begin{figure}
  \centering
  \includegraphics[width=0.95\linewidth, trim=0 1.5cm 0 0, clip]{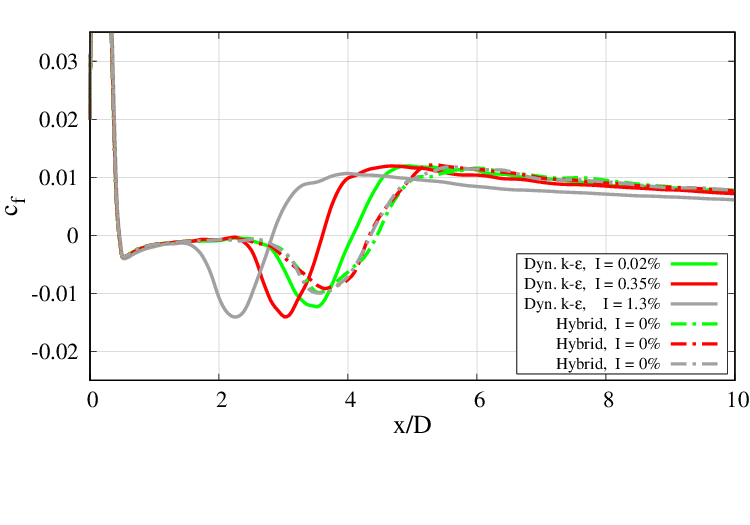}\mylab{0mm}{40mm}{(\aaa)}
  \includegraphics[width=0.95\linewidth, trim=0 1.5cm 0 0, clip]{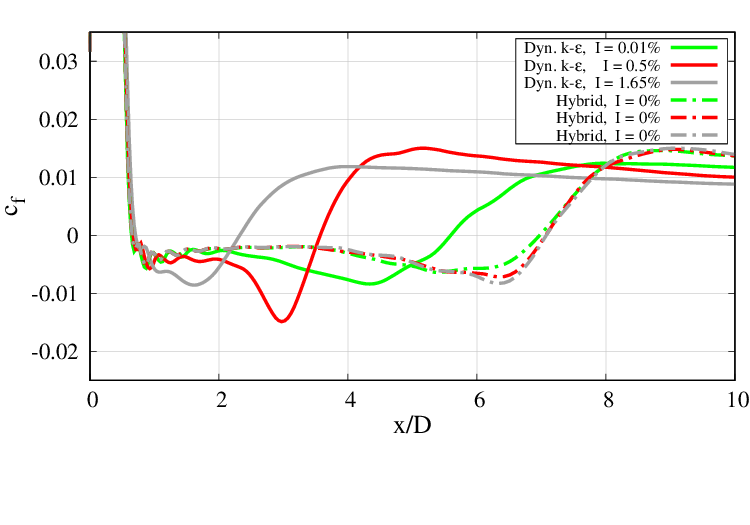}\mylab{0mm}{40mm}{(\bbb)}
  \caption{Average friction coefficient for the scale-resolving 3D unsteady simulations performed in (a) the fine and (b) coarse grids of the new dynamic RANS closure compared to the hybrid LES/RANS approach.}
\label{fig_friction_3}
\end{figure}

The developed dynamic procedure has shown to be able to capture relevant unsteady and 3D coherent structures of turbulence and transitional phenomena in separating and reattaching flows. Accordingly, it appears that the dynamic RANS model enables classical eddy viscosity models to capture relevant turbulent or transitional flow features when the grid resolution allows to represent them. This is a typical feature of scale-resolving methods that is why we compare also our results with those of the hybrid LES/RANS model. As shown in figure \ref{fig_friction_3}(a), the shape of the friction coefficient profile for the hybrid LES/RANS approach highlights that the physics of laminar separation, transition and turbulent reattachment is reproduced as much as that of the new dynamic RANS closure. The hybrid model is, however, found to be unresponsive to changes in the boundary conditions for $k$ and $\varepsilon$. This lack of sensitivity is caused by the LES-like mode that is operated in the region of the domain between the inlet and the solid body. As a result, the free-stream turbulence intensity cannot be imposed through boundary conditions on $k$ and $\varepsilon$, but should instead be generated through a forcing plane and then directly resolved as it is in all scale-resolving methods. In the present simulations such a forcing plane is not implemented and, as a result, the free-stream turbulence intensity that reaches the plate is extremely low (virtually zero) regardless of the imposed inlet boundary conditions. This makes the present hybrid LES/RANS approach incapable of correctly modeling the effect of free-stream turbulence intensity through boundary conditions contrary to the present dynamic RANS approach. This is a key aspect that makes the present dynamic RANS procedure definitely favorable for the numerical solution of complex flows commonly encountered in applications. 

Finally, let us point out that when coarsening the spatial resolution, the resolved physics deteriorates more for the hybrid LES/RANS approach than for the present dynamic RANS model, see figure \ref{fig_friction_3}(b). As already pointed out when analysing the instantaneous flow pattern reported in figure \ref{fig_inst_fields_coarse}, the quality of the solution increasingly relies on the ability of the closure in capturing the physics of the regions calling for a modeled stress increase as they become larger when coarsening the resolution. The developed dynamic RANS approach is found to outperform in transitional regions where scale-resolving features are partially supported by the grid and in the fully modeled regions by improving the flow solution with respect to classical pure RANS closures as it will be demonstrated in \S\ref{sec_2D_results} through the analysis of 2D steady simulations. This is a fundamental feature rendering the present dynamic formulation more suitable for real-world CFD applications where the grid resolution assigns the task of solving the flow to statistical closures in large portions of the domain.

\subsection{Mean velocity and turbulent profiles}

\begin{figure*}
  \centering
  \includegraphics[width=0.48\linewidth, trim=0 1.4cm 0 0, clip]{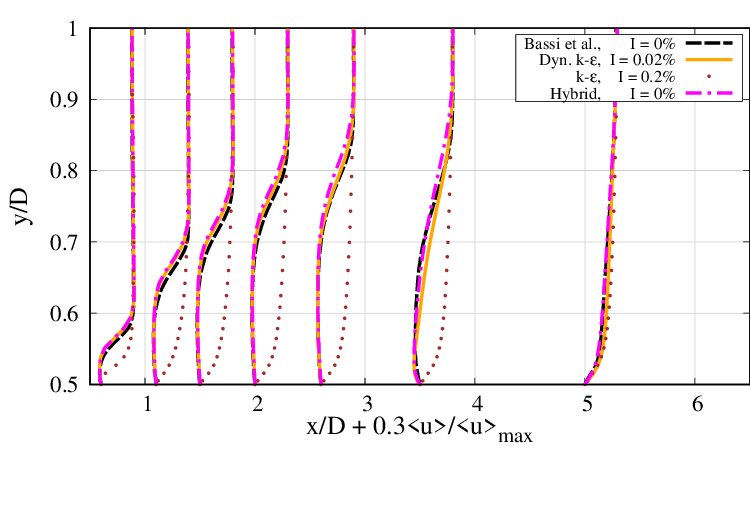}\mylab{-86mm}{40mm}{(\aaa)}
  \includegraphics[width=0.48\linewidth, trim=0 1.4cm 0 0, clip]{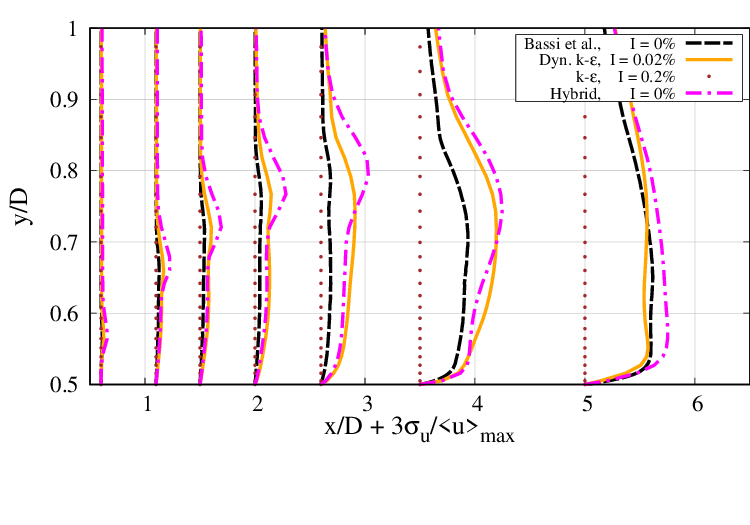}\mylab{1mm}{40mm}{(\bbb)}
    \includegraphics[width=0.48\linewidth, trim=0 1.4cm 0 0, clip]{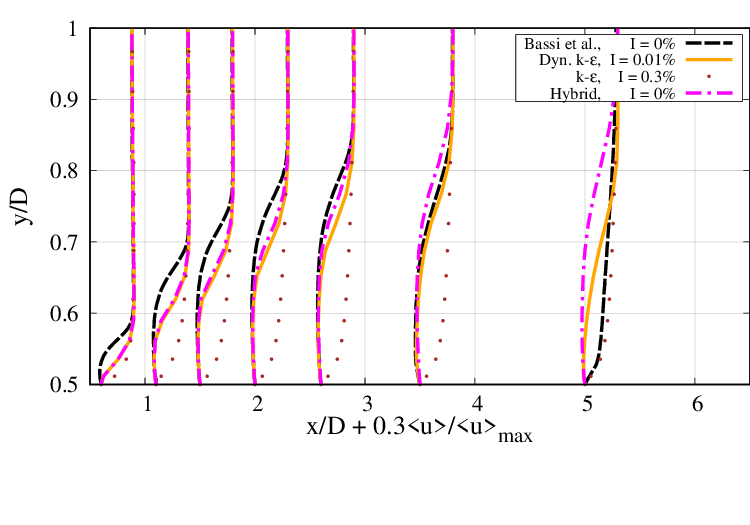}\mylab{-86mm}{40mm}{(\ccc)}
  \includegraphics[width=0.48\linewidth, trim=0 1.4cm 0 0, clip]{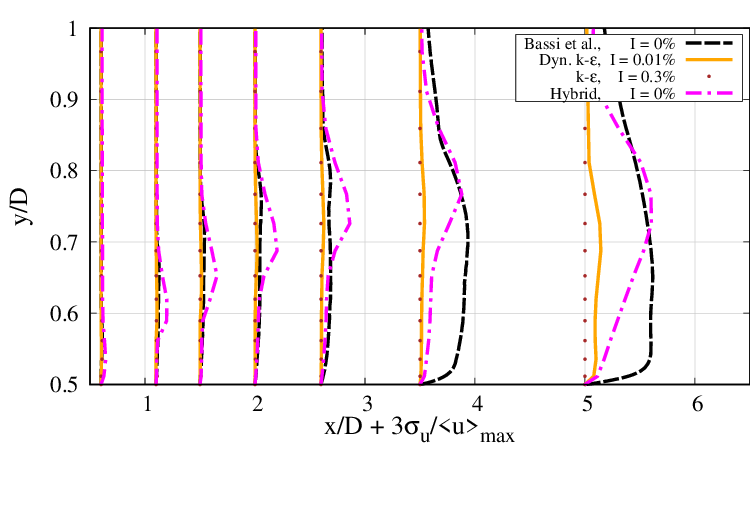}\mylab{1mm}{40mm}{(\ddd)}
  \caption{Mean streamwise velocity profiles (a) and (c) and mean streamwise turbulent intensities (b) and (d) at the lowest free-stream turbulence level. Results from scale-resolving simulations performed on the fine grid (a) and (b) and on the coarse grid (c) and (d). The dynamic RANS formalism, the $k-\varepsilon$ model and the hybrid LES/RANS approach are compared with the high-fidelity implicit-LES simulations performed by \citet{Crivellini20}.}
\label{fig_traverse}
\end{figure*}

To further assess the quality of the transitional flow solutions over a rounded leading-edge flat plate achieved by the different modeling approaches, we address here the wall-normal profiles of the mean streamwise velocity $\langle u \rangle$ and of the standard deviation of streamwise velocity $
\sigma_u = \sqrt{\langle u'u'\rangle}$ at different streamwise locations. As shown in figure \ref{fig_traverse}(a) and (b), the dynamic RANS and hybrid LES/RANS closures show a similar behaviour on the finer grid, well reproducing the scaling of the mean velocity and turbulent profiles of the reference implicit-LES simulation \citep{Crivellini20}. The main discrepancy is an overestimation of the intensity of turbulent fluctuations along the leading-edge shear layer developing in the initial part of the separating bubble. However, the overall phenomena of laminar separation, transition and turbulent reattachment are found to be nicely captured by the two modeling approaches in accordance with the analysis of the previous sections. On the contrary, the baseline $k-\varepsilon$ model completely fails in reproducing the flow separation thus confirming again its inadequacy in capturing laminar boundary layer separation.

When coarsening the grid resolution, the accuracy of the flow phenomena deteriorates as shown in figure \ref{fig_traverse}(c) and (d). In particular, a significantly longer flow recirculation is reproduced by both the dynamic RANS and hybrid LES/RANS closures as already shown when analysing the friction coefficient in figure \ref{fig_friction_3}(b). As a consequence, the mean standard deviation profiles of streamwise velocity are significantly misaligned with respect to those of the reference implicit-LES solution. In any case, the solution exhibits a slightly better behavior for the dynamic RANS approach than for the hybrid LES/RANS approach. The solution of the baseline $k-\varepsilon$ model is found to be less sensitive to the coarsening of the mesh but still completely failing in reproducing the flow physics.


\section{Statistical approach: 2D steady simulations}
\label{sec_2D_results}

One of the distinctive features of the proposed dynamic formalism is its ability to automatically switch from scale-resolving to statistical approaches depending on the numerical setup employed. This is a key property for applications where statistical closures are still favoured because of the ease of use associated with their very low computational costs. Indeed, RANS statistical approaches seek the average flow solution, which in many applications exhibits symmetries such as steadiness in time and homogeneity in some spatial directions, allowing for drastic reductions of the computational costs by avoiding the need for time integration and by reducing the dimensionality of the computational domain.

In order to explore the potential of the proposed dynamic RANS closure also in statistical approaches, we report here results of the T3L flow case simulated in steady-state and 2D conditions. This setup is not suitable for scale-resolving methods such as the hybrid LES/RANS approach that is why results of the proposed dynamic RANS approach will be compared solely with those of the baseline $k-\varepsilon$ model together with the high-fidelity implicit-LES simulations performed by \citet{Crivellini20} as a reference. Statistics are obtained by averaging over the last 900 iterations of the simulations. This value was selected based on standard statistical convergence criteria, taking into account the behaviour of the friction coefficient profile as the statistical sample size varies. The specific settings of the dynamic procedure have been already reported in section \S\ref{sec_dyn_settings} but we remind here that for steady-state simulations we adopt an accumulation between successive iterations (\ref{eq_accum_ave}) as test filter operator.

\begin{figure}
  \centerline{\includegraphics[width=0.95\linewidth, trim=0 1.5cm 0 0, clip]{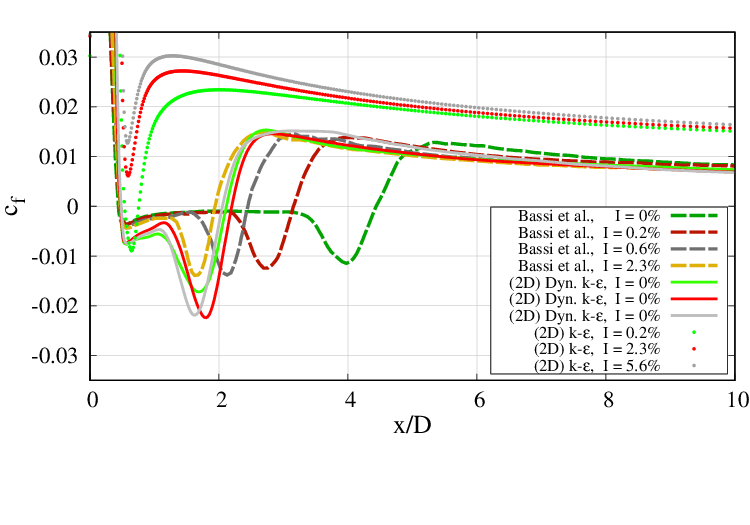}}
  \caption{Average friction coefficient for the statistical 2D steady simulations performed with the new dynamic RANS formalism and with the standard $k-\varepsilon$ model. Results are compared with the high-fidelity implicit-LES simulations performed by \citet{Crivellini20}.}
\label{fig_2Dcf}
\end{figure}

The behaviour of the friction coefficient $c_f$ is reported in figure \ref{fig_2Dcf}. The friction coefficient profiles resulting from the 2D dynamic RANS simulations reveal substantial differences with respect to those obtained through 3D simulations. They consist in a overestimation of the negative peak of $c_f$  within the reverse boundary layer and in the loss of responsiveness to the inlet boundary conditions meant to model the free-stream turbulence intensity. The latter is essentially induced by the fact that the dynamic procedure sets the model coefficient $c_\mu$ to extremely small values in the upstream flow region. Hence, the free-steam turbulence level perceived by the body is always virtually zero, regardless of the free-stream conditions imposed through the inlet boundary conditions of $k$ and $\epsilon$. Despite these deficiencies, the dynamic RANS simulations still yield a distinct reproduction of the main features of the flow and a correct evaluation of the intensity of the $c_f$ along the downstream development of the attached boundary layer. The quality of these features becomes apparent when comparing these results to those obtained with the baseline $k-\varepsilon$ model. Indeed, the $k-\varepsilon$ model does not capture transition thus drastically reducing or preventing separation. The friction coefficient is then largely overestimated thus affecting the entire downstream development of the flow. In particular, the values of the friction coefficient are found to remain overestimated also along the downstream evolution of the attached boundary layer. Overall, these results show that the developed dynamic formalism leads to a significant improvement of the quality of the solution also in statistical approaches. This is a fundamental property for most of CFD applications and explains the improved performances that the dynamic RANS closure exhibits also in scale-resolving approaches when coarsening the grid, as already observed in section \S\ref{sec_3D_results}.


\section{Conclusions}
\label{sec_concl}

It is well known that numerical solutions of the equations of motion adopting statistical turbulence closures predict unsteady flow patterns in certain flow problems where the average solution is on the contrary steady. It is also well known that the combination of statistical and scale-resolving closures, the so-called hybrid approaches, allows to capture the turbulent structure of the solution in many flow problems. The numerical approaches provided by both statistical and hybrid closures are widely used in CFD applications but lack a clear theoretical framework for their assessment and improvement. From one side, the formalism provided by the exact RANS equations that is based on the average operator does not predict unsteady nor scale-resolving flow solutions. From the other side, the formalism provided by the exact LES equations that is based on a spatial filtering operator does not predict the loss of scale-resolving features in the flow solution. These simple facts suggest that an alternative theoretical framework must be used to assess and eventually develop turbulence closures when employed in very-coarse steady and unsteady numerical settings typical of CFD applications.

\subsection{A change of paradigm: temporally filtered Navier-Stokes equations}

In the present work we propose an alternative operatorial approach for the development of an exact theoretical framework that is able to describe the smoothing action of turbulence closures in both statistical and scale-resolving approaches. The basic assumption is that such a smoothing action can be described by a temporal filtering operator. As a consequence, the proper exact theoretical framework for turbulence closures ranging from the statistical to the scale-resolving ones, is represented by the temporally filtered Navier-Stokes equations. Indeed, the relevant aspect of this exact formalism is given by the fact that the related turbulent stresses $R(u_i, u_j) = \langle u_i u_j \rangle_\tau - \langle u_i \rangle_\tau \langle u_j \rangle_\tau$ smoothly converge from those related to scale-resolving methods (subgrid stresses) to those related to statistical approaches (Reynolds stresses) by increasing the filter time scale $\tau$. In other words, the temporally filtered Navier-Stokes equations represent an unifying exact theoretical framework that provide a direct link between the scale-resolving and statistical approaches. In particular, scale-resolving approaches are interpreted as the result of a turbulence closure whose additional stresses are associated to a filter time scale that is not large enough to recover the averaged solution, i.e. $\tau U/L \le 1$ and $\langle \cdot \rangle_\tau \neq \langle \cdot \rangle$. On the other hand, statistical approaches are interpreted as the result of a turbulence closure whose additional stresses are associated to a very large filter time scale, i.e. $\tau U/L \gg 1$ and $\langle \cdot \rangle_\tau = \langle \cdot \rangle$. The temporally filtered Navier-Stokes equations and their link between the statistical and scale-resolving approaches have already attracted some interest in scientific community \citep{pruett2003temporally, pruett2008temporal, fadai2010temporal, friess2015toward, duffal2022development}. The potential of this framework is highlighted here by unveiling its relevant fundamental properties and by deriving a new class of turbulence closures able to bridge between scale-resolving and statistical approaches.

\subsection{Dynamic RANS closures}

Beyond providing an exact theoretical framework for actual CFD implementations, the proposed change of paradigm enables a very important algebraic property that can be leveraged to improve the predictive capabilities of RANS eddy viscosity models. It consists in a exact relation between turbulent stresses at different temporal filter levels that we use to develop a dynamic procedure for the calculation of the model coefficient $c_\mu$ appearing in the definition of eddy viscosity that for a standard $k-\varepsilon$ approach reads $\nu_\tau = c_\mu k_\tau^2 / \epsilon_\tau$. The results highlight that the developed dynamic RANS closure shows accurate scale-resolving properties when the numerical settings and the grid resolution allow them to develop. In particular, the dynamic RANS model is found to capture challenging flow features such as the laminar to turbulence transition and the dependency of the boundary layer separation and reattachment to changes in the free-stream turbulence applied as boundary condition. These are two key properties of overwhelming relevance for CFD applications where separating and reattaching phenomena are often present whose features are heavily dependent on transition and free-stream turbulence. The dynamic RANS formalism is found to outperform also when the grid resolution is coarsened. The reason is the improved physics captured by the dynamic RANS closure also when employed in statistical approaches. Indeed, the regions of the flow that call for a modeled stress increase typical of statistical approaches, widen with the coarsening of the grid. As a consequence, the quality of the entire flow solution increasingly relies on the quality of the solution in the fully modeled regions. Tests of the dynamic RANS closure when employed as a statistical closure in 2D steady-state numerical settings, reveal a significant improvement of the flow physics captured especially in the laminar to turbulence transitional regions where the tendency of the baseline $k-\varepsilon$ approach to over-predict eddy viscosity is avoided. As a result, the dynamic RANS formalism is found to be robust to a coarsening of the spatial resolution in scale-resolving approaches.

\subsection{Perspectives}

Overall, the improved performances in scale-resolving unsteady approaches (in terms of ability of capturing laminar to turbulence transition, of responsiveness to the free-stream turbulence levels applied through the boundary conditions, and of robustness to a coarsening of the spatial resolution) and the improvement of the physics of the solution captured in statistical steady-state approaches, render the developed dynamic RANS closure promising for actual CFD applications. This result sets the stage for further more technical studies that are deemed essential to investigate alternative model implementations and to address its performance over a wider range of applications and numerical conditions. Regarding the former, it is important to highlight that the developed dynamic RANS closure actually represents a general formalism that forms a class of turbulence closures where the dynamic procedure can be applied to different eddy viscosity models as base closure. Hence, the present implementation based on the standard $k-\varepsilon$ closure is just one version over many possible and can be reasonably improved by adopting more advanced closures such as the $k-\omega$ and $k-\omega$ SST models. Hence, further studies addressing the best base turbulence closure for the dynamic RANS formalism are envisaged and left to future more technical works. Regarding the latter, it is important to note that the most critical aspect for hybrid LES/RANS approaches is given by the strict temporal resolution constraint required by the scale-resolving part of the modelling approach usually provided by classical LES models. The developed dynamic RANS approach is a RANS model in nature where scale-resolving features are enabled by the dynamic procedure. Hence, we might expect a robustness to a coarsening of the temporal resolution typical of classical RANS modelling approaches. Hence, further studies assessing the sensitivity of the flow solution to the temporal resolution are also envisaged in order to gain relevance in widespread industrial contexts.

\subsection{Additional remarks}

In addition to serving as a unifying formalism for scale-resolving and statistical approaches, the exact framework provided by the temporally filtered Navier-Stokes equations exhibits two relevant properties that are remarked here. First, the commutative property of the temporal filtering operator with respect to differential operators can be recovered by real CFD applications more easily than the commutative property of the spatial filtering operator. Indeed, by assuming the length and time scales of the filtering operator as imposed by the spatial and temporal discretization scales of the simulation, $\Delta = \sqrt[3]{\Omega}$ and $\tau = \Delta t$, where $\Omega$ and $\Delta t$ are the volume of the grid elements and the time step, the temporal filtering commutation error can easily be avoided by performing simulations at a constant time step. In contrast, the need to resolve spatially inhomogeneous flow features demands the use of inhomogeneous grid resolutions, thus making the spatial filtering operator generally non-commutative with differential operators \citep{klein2020analysis}. The second relevant property of the temporally filtered Navier-Stokes equations is related to the use of explicit filtering operator such as those employed here in the dynamic RANS approach. Indeed, an explicit filter kernel can be applied to all the flow regions without ambiguity in the temporal filtering formalism, contrary to the spatially filtered one typical of LES where the definition of its kernel is challenging because of issues related to boundary constraints imposed by the problem geometry and to the stretching of the grid.

Finally, let us remark a basic aspect of the dynamic RANS closure here developed. The model definition does not explicitly involves the temporal and spatial resolution employed for the numerical solution of the equations of motion. Such information is usually explicitly taken into account in scale-resolving approaches, the most famous example is the Smagorinsky model for LES where the grid resolution is explicitly included in the definition of the eddy viscosity through the filter length, $\Delta = \sqrt[3]{\Omega}$ where $\Omega$ is the volume of grid elements. On the contrary, such information is not explicitly included in statistical closures because their solution is independent on the numerical resolution by definition, the most famous example are the two-equation eddy viscosity models for RANS. The present dynamic RANS closure is developed to span scale-resolving and full statistical approaches. Hence, an explicit dependence on the numerical resolution on the model definition is not adopted on purpose in order to address also statistical approaches. On the other hand, the dependence on the spatial and temporal resolution is implicitly recovered by the dynamic procedure itself when addressing problems with a scale-resolving approach. In particular, the dynamic RANS procedure is based on a measure of the change of the solution among two time scales through the use of the test filter (\ref{eq_moving_ave}). The spatial resolution determines the type of turbulence structures that can be generated and that span these two time scales. Eventually, the dynamic procedure adapts the eddy viscosity through equation (\ref{eq_dyn_eq}) to not dissipate them. In other words, the spatial and temporal filtering implicitly performed by the dynamic eddy viscosity when employed in scale-resolving approaches is strictly dependent on the numerical resolution adopted although not explicitly taken into account in the model definition.


\begin{acknowledgments}
We wish to acknowledge professors Antonella Abb\'a and Massimo Germano for the fruitful discussions on the preliminary draft of the present work.
\end{acknowledgments}


\section*{Data Availability Statement}

The data that support the findings of this study are available from the corresponding author upon reasonable request.

\appendix


\section{Flow solutions around a rectangular cylinder}
\label{app_barc}

To further assess the performances of the developed dynamic RANS approach, the flow around a rectangular cylinder with aspect ratio 5 is also simulated. This flow case is the object of an international initiative known as BARC (Benchmark on the Aerodynamics of a Rectangular 5:1 Cylinder) \citep{Bruno} and several numerical studies have been conducted\citep{BARC_Mannini, wornom2011variational, cimarelli2018structure, BARC_ChiariniQuadrio, BARC_Crivellini}. Among others, a Direct Numerical Simulation (DNS) at a fairly high Reynolds number $Re = U_0 D / \nu = 14000$ has been recently reported by \citet{cimarelli2024reynolds}. Here, $U_0$ is the free-stream velocity and $D$ is the thickness of the rectangular cylinder. The distinguishing feature of the BARC case compared to the T3L case, is the laminar boundary layer separation that is fixed at the sharp leading-edges of the body.

\begin{figure}
  \centering
  \includegraphics[width=0.48\linewidth, trim=0 1.cm 0 0, clip]{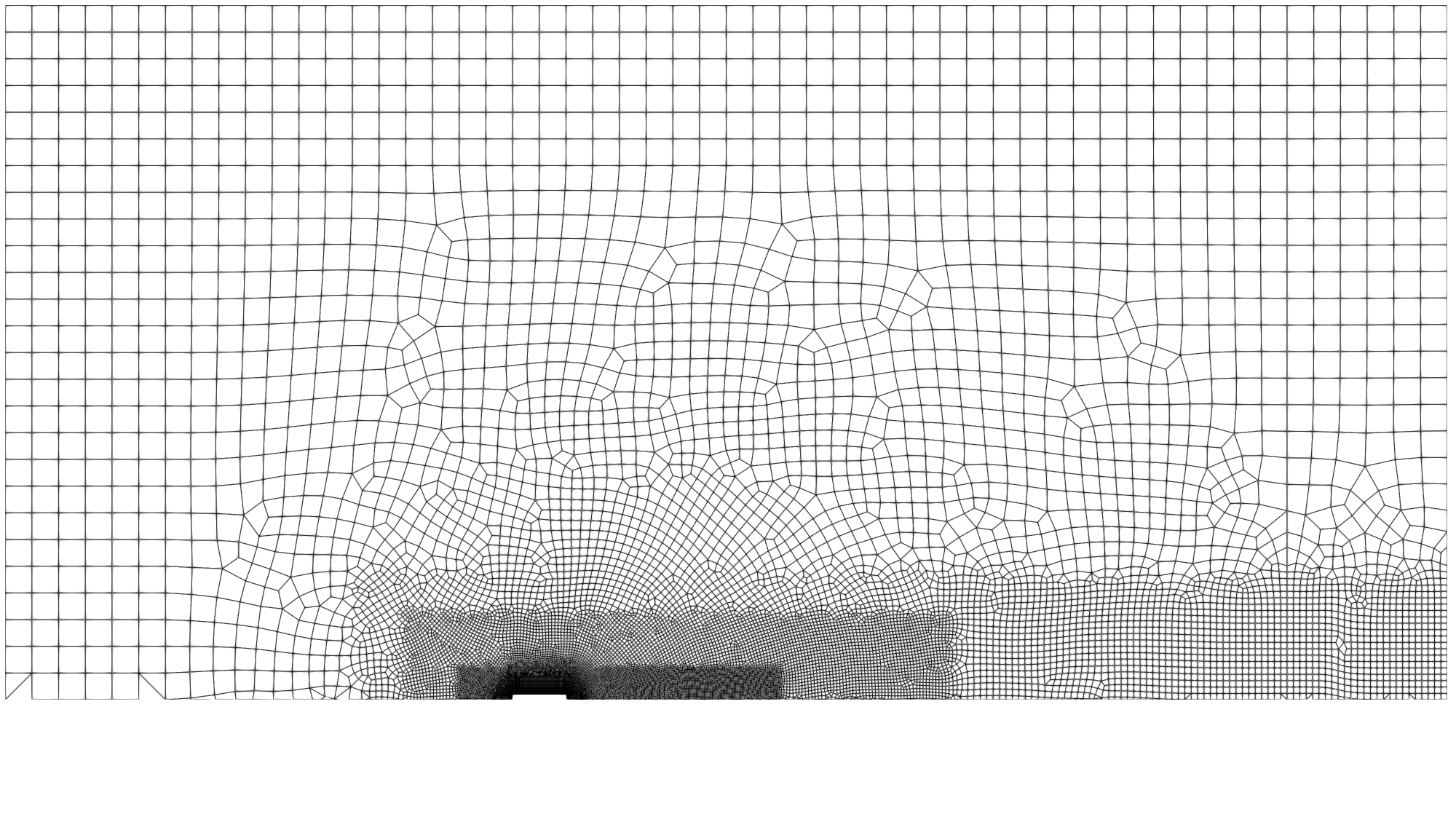}
  \includegraphics[width=0.48\linewidth, trim=0 1.cm 0 0, clip]{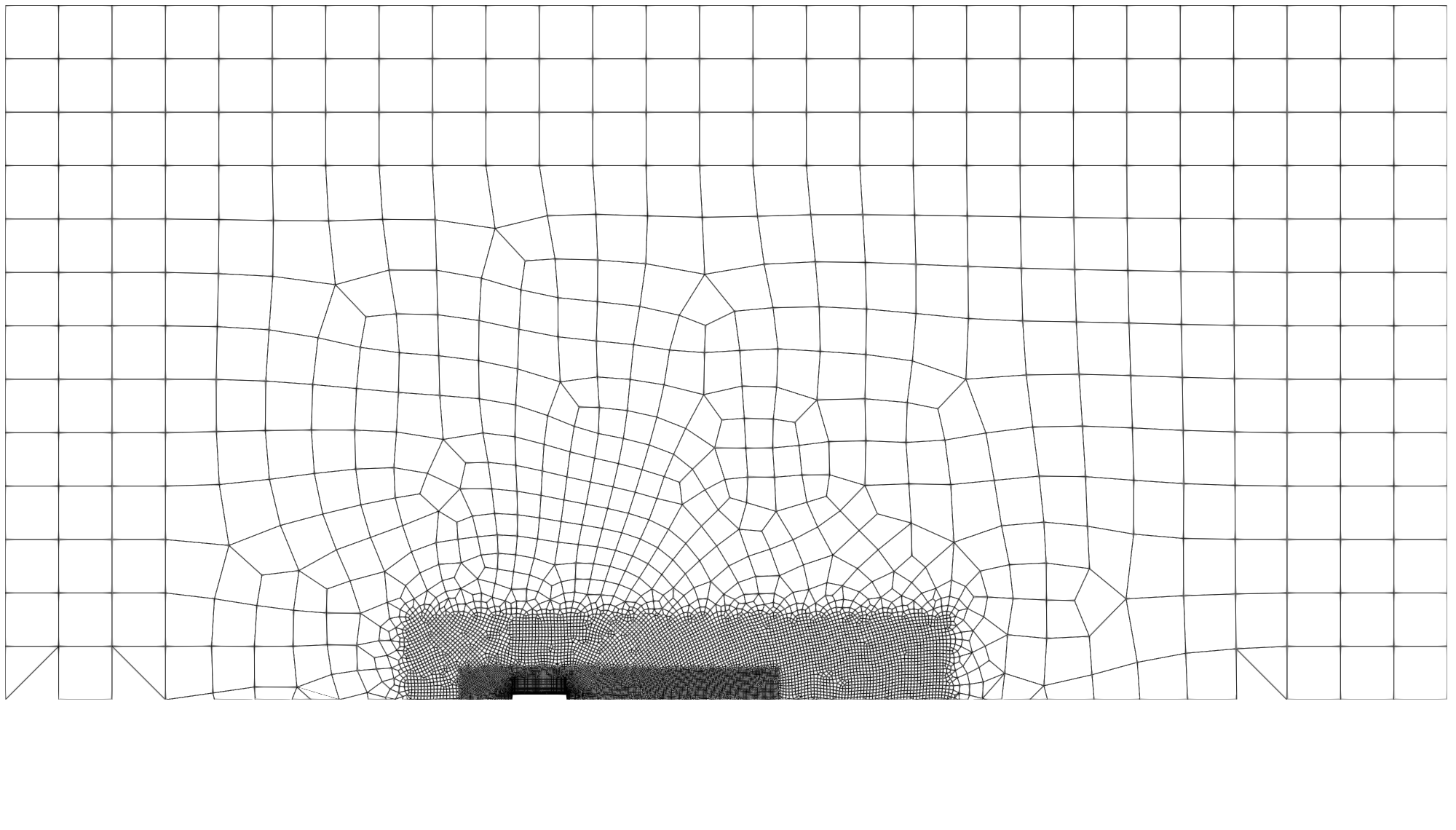}
  \caption{Fine (left) and coarse (right) grids developed for the simulation of the flow around a rectangular cylinder with aspect ratio 5.}
\label{fig_FMesh_BARC}
\end{figure}

The numerical schemes and the dynamic procedure settings are those used for the simulations of the T3L case and reported in sections \S\ref{subsec:NC} and \S\ref{sec_dyn_settings}. A constant time step is adopted in order to obtain a condition $CFL \approx 1.5$. In order to study the effect of the grid resolution, two grids have been developed. Both grids are stretched in order to increase the spatial resolution by moving towards the plate in the $y$-direction and towards the leading- and trailing-edges in the $x$-direction. The finer grid is composed by $2.48M$ mesh elements and is shown in figure \ref{fig_FMesh_BARC}. The resulting spatial resolution evaluated in a refined box surrounding the rectangular cylinder up to a distance $1.5D$, is $\Delta x/D = 0.027 \div 0.05$, $\Delta y/D = 0.0028 \div 0.068$ and $\Delta z/D = 0.1$. On the other hand, the coarser grid has been obtained by roughly halving the number of cells in the spanwise $z$-direction and in the $(x-y)$-plane, thus leading to a total number of $0.92M$ grid elements, see again figure \ref{fig_FMesh_BARC}. The resulting spatial resolution evaluated again in a box surrounding the rectangular cylinder up to a distance $1.5D$, is $\Delta x/D = 0.023 \div 0.15$, $\Delta y/D = 0.0028 \div 0.11$ and $\Delta z/D = 0.14$. A free-stream velocity $U_0$ is imposed at the inlet and at the top and bottom boundaries. A convective boundary condition is applied at the outlet and periodic boundary conditions are imposed in the spanwise direction. Finally, statistics are computed by averaging in the spanwise direction and over 140 time samples collected every $5D/U_0$ times once the flow has reached a fully developed state. The resulting average operator is hereafter denoted as $\langle \cdot \rangle$.

\begin{figure*}
  \centering
  \includegraphics[width=0.45\linewidth]{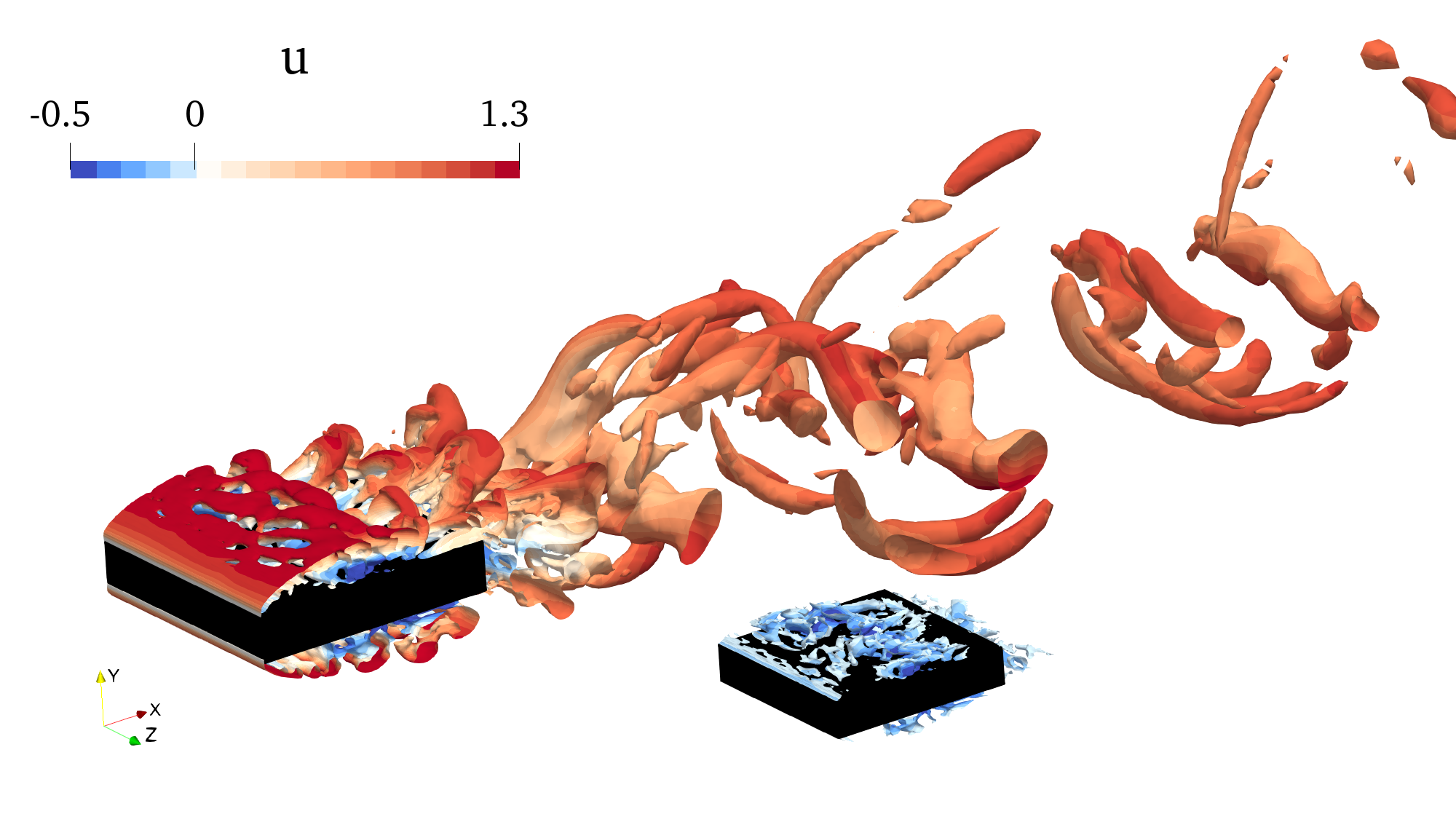}\mylab{-80mm}{28mm}{(\aaa)}
  \hspace{0.5cm}\includegraphics[width=0.45\linewidth]{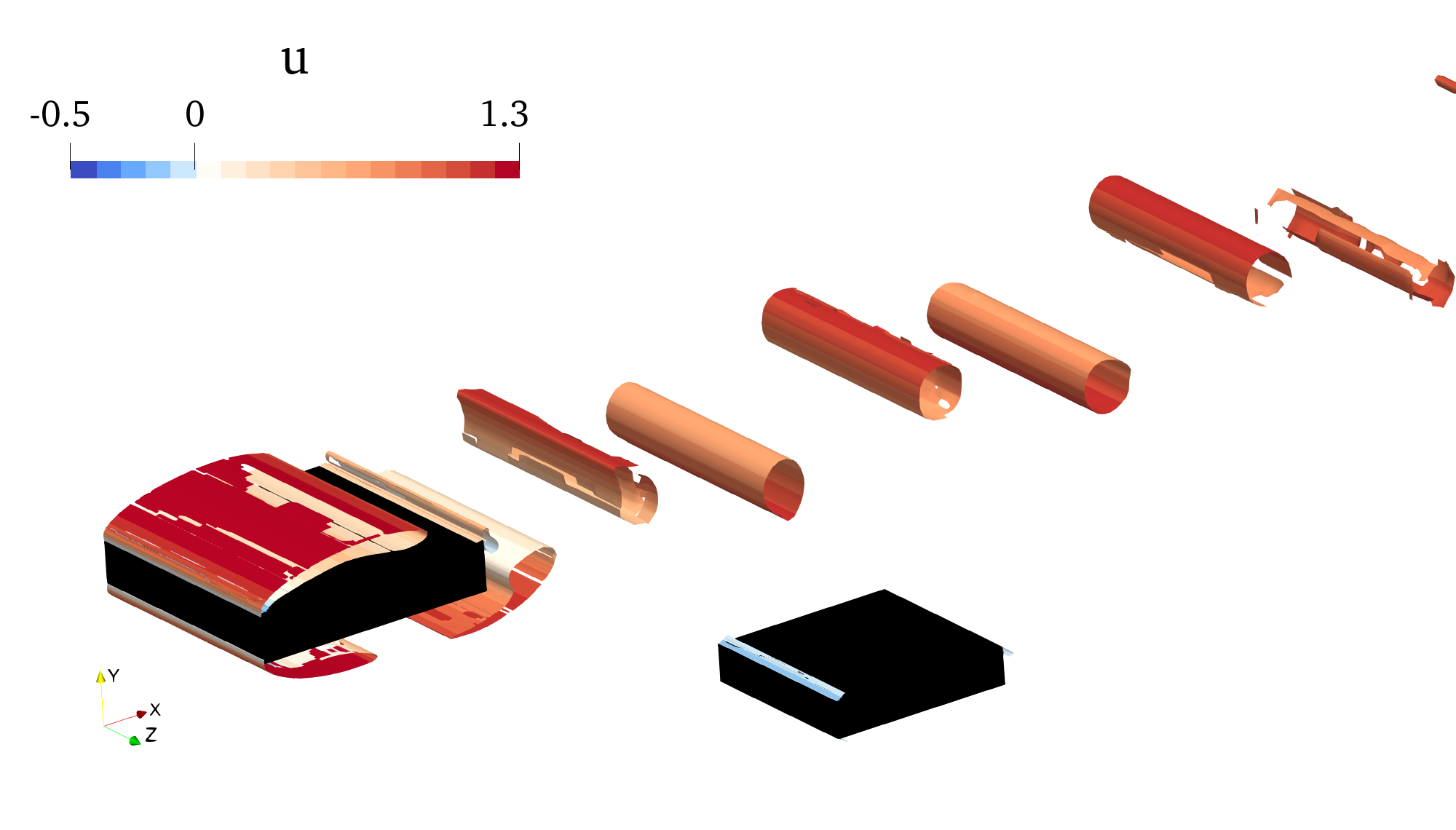}\mylab{-80mm}{28mm}{(\bbb)}
  \includegraphics[width=0.45\linewidth]{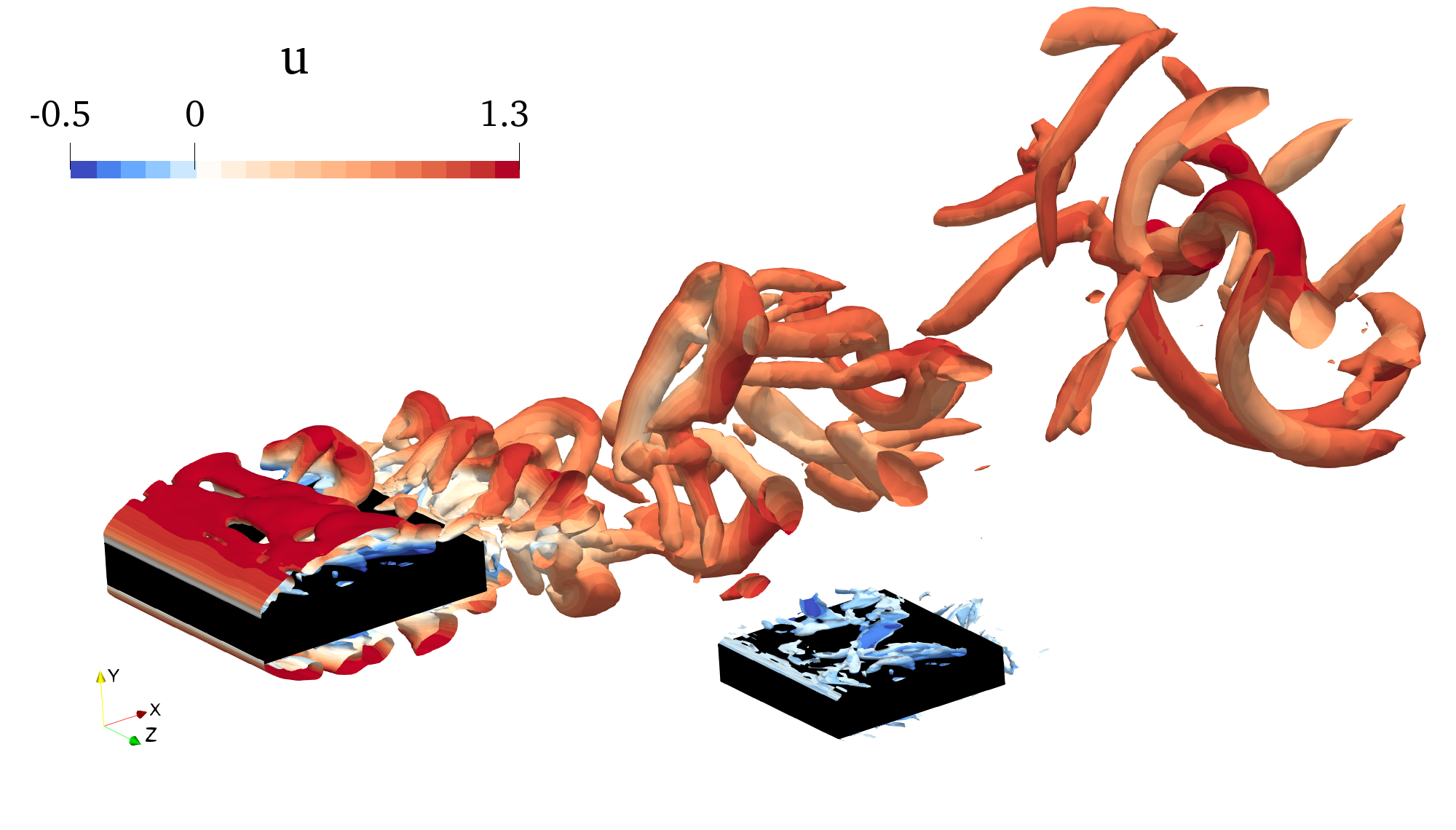}\mylab{-80mm}{28mm}{(\ccc)}
  \hspace{0.5cm}\includegraphics[width=0.45\linewidth]{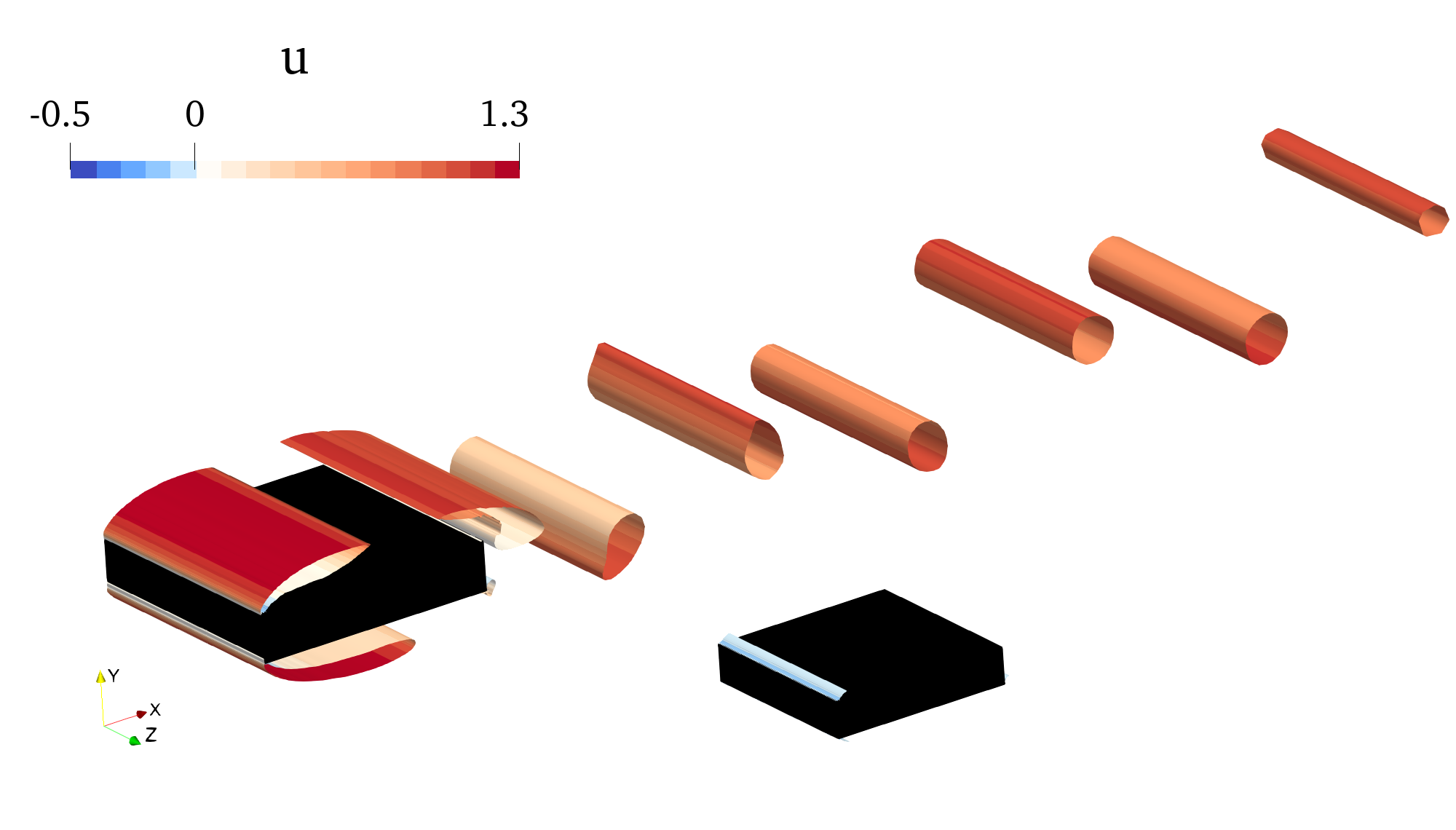}\mylab{-80mm}{28mm}{(\ddd)}
  \caption{Separating and reattaching flow around a rectangular cylinder. Instantaneous flow realizations reported with iso-surfaces the Q-Criterion \citep{hunt1988eddies}, $Q = 0.15$, colored with streamwise velocity. In the inset panels, isosurfaces with positive streamwise velocity are hidden to reveal the structures of the reverse boundary layer within the main recirculating bubble. Flow solutions obtained with the dynamic RANS approach are reported in (a) and (c) for the finer and coarser grids respectively. Flow solutions obtained with the standard $k-\varepsilon$ implementation are reported in (b) and (d) for the finer and coarser grids respectively.}
\label{fig_BARC_inst}
\end{figure*}

The instantaneous flow solutions are reported in figure \ref{fig_BARC_inst}. Analogously to the T3L flow case analysed in the main body of the present work, the solution obtained with the dynamic RANS procedure is found to capture 3D and unsteady flow structures typical of the flow around rectangular cylinders \citep{cimarelli2018structure}, see figure \ref{fig_BARC_inst}(a). Furthermore, the dynamic RANS solution is also capable of spawning and transporting turbulent structures in the reverse boundary layer within the main recirculating bubble. When coarsening the spatial resolution, the dynamic RANS solution is found to maintain these flow features thus highlighting a certain degree of robustness of the approach, see figure \ref{fig_BARC_inst}(c). Indeed, the only difference is the obvious lack of fine scale structures in the coarser grid solution. Contrary to the dynamic RANS, the standard $k-\varepsilon$ implementation completely misses the phenomenological features of the separating and reattaching flow by reproducing a flow solution composed by 2D, albeit unsteady, structures, see figures \ref{fig_BARC_inst}(b) and (d).

\begin{table}
  \begin{center}
\def~{\hphantom{0}}
  \begin{tabular}{cccc}
                                     &  $c_d$ & $\sigma_{c_l}$ & $St$ \\[3pt]
       DNS (Cimarelli et al.)        &  1.038 &  0.63 & 0.111\\
       Dyn. $k-\varepsilon$ (fine)   &  1.029 &  0.69 & 0.116\\
       Dyn. $k-\varepsilon$ (coarse) &  1.080 &  0.92 & 0.112\\
       $k-\varepsilon$ (fine)        &  1.064 &  0.66 & 0.122\\
       $k-\varepsilon$ (coarse)      &  1.056 &  0.54 & 0.118
  \end{tabular}
  \caption{Average drag coefficient $c_d$, standard deviation of the lift coefficient $\sigma_{c_l}$ and Strouhal number of the vortex shedding $St = f U_0/D$ measured by using the frequency $f$ of the peak of the lift coefficient spectrum.}
  \label{tab_TIQ}
  \end{center}
\end{table}

\begin{figure*}
  \centering
  \includegraphics[width=\linewidth]{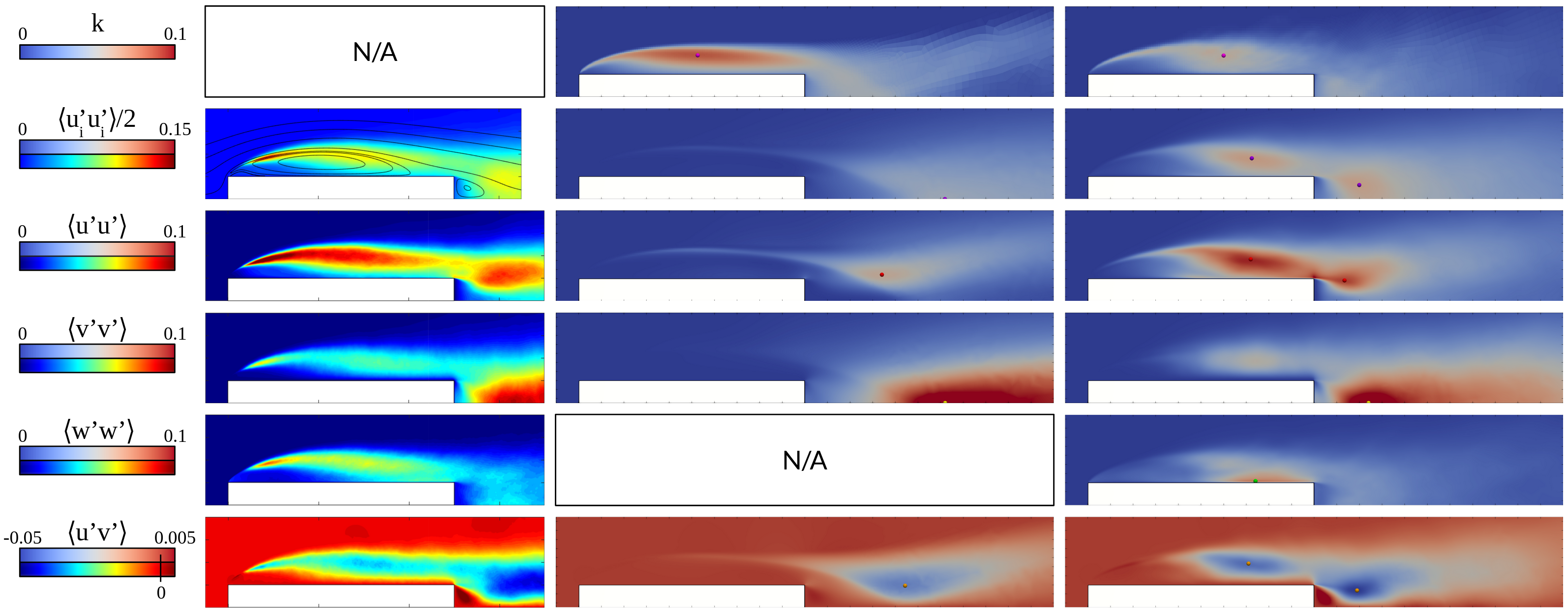}\mylab{-140mm}{70mm}{DNS}\mylab{-92mm}{70mm}{$k-\varepsilon$}\mylab{-36mm}{70mm}{Dyn. $k-\varepsilon$}
  \caption{Turbulent kinetic energy and Reynolds stresses in the flow around a rectangular cylinder. The left panels report the behaviour obtained from DNS \citep{cimarelli2024reynolds} while the central and right panels show the solution from the standard $k-\varepsilon$ and dynamic RANS approaches, respectively.}
\label{fig_TKER}
\end{figure*}

From a statistical point of view, the quality of the solution is addressed first by evaluating the predictions of the aerodynamic forces, in particular the drag coefficient, the standard deviation of the lift coefficient and the Strouhal number of vortex shedding. As shown in table \ref{tab_TIQ}, the dynamic RANS formalism nicely reproduces such quantities when compared with DNS data from \citet{cimarelli2024reynolds} with a slight deterioration of the estimates with the coarsening of the grid. Despite the poorly reproduced physics, these integral quantities are also well reproduced by the standard $k-\varepsilon$ model.

The behaviour of turbulent kinetic energy and of Reynolds stresses for the dynamic RANS and standard $k-\varepsilon$ models is reported in figure \ref{fig_TKER} compared with results from DNS data \citep{cimarelli2024reynolds}. When considering turbulent kinetic energy, both the resolved and modelled contributions are reported, $\langle u_i'u_i'\rangle /2$ and $k$ respectively. Only the results from the finer grid simulations are reported for brevity and because the results from the coarser grid are not too different. The difference between the dynamic RANS and standard $k-\varepsilon$ solutions is marked. As already pointed out, the standard $k-\varepsilon$ model introduces a high amount of eddy viscosity thus damping all turbulent fluctuations except to a shedding of two-dimensional vortices. As a result, the resolved turbulent kinetic energy is very weak compared to the modelled one. For the same reason, the Reynolds stresses are much lower in intensity with respect to those predicted by the reference DNS. The only exception are the cross-stream fluctuations $\langle v'v'\rangle$ in the near wake as a consequence of the higher intensity associated to the shedding of 2D vortices. On the contrary, the dynamic RANS formalism adjusts the eddy viscosity to be compliant with the spatial resolution available, thus allowing for the proper development of a wide range of turbulent structures. As a matter of facts, the resolved turbulent kinetic energy and Reynolds stresses reproduce those of the reference DNS quite well in both shape and intensity.


\section{Turbulent flow over a bump}
\label{app_bump}

As a final assessment of the dynamic RANS formalism, we address here its solution in the turbulent flow over a bump. The flow settings consist in a fully turbulent channel with a bump in the bottom wall that produces a turbulent flow separation. Hence, the main difference with respect to the previous flow settings is the absence of laminar to turbulent transitional phenomena. The specific flow settings are those corresponding to the flow case A3 addressed through DNS by \citet{mollicone2017effect}. It corresponds to a bulk Reynolds number $Re_b = U_b h / \nu = 10000$, where $U_b$ is the bulk velocity and $h$ is the channel half height.

\begin{figure}
  \centering
  \includegraphics[width=0.95\linewidth]{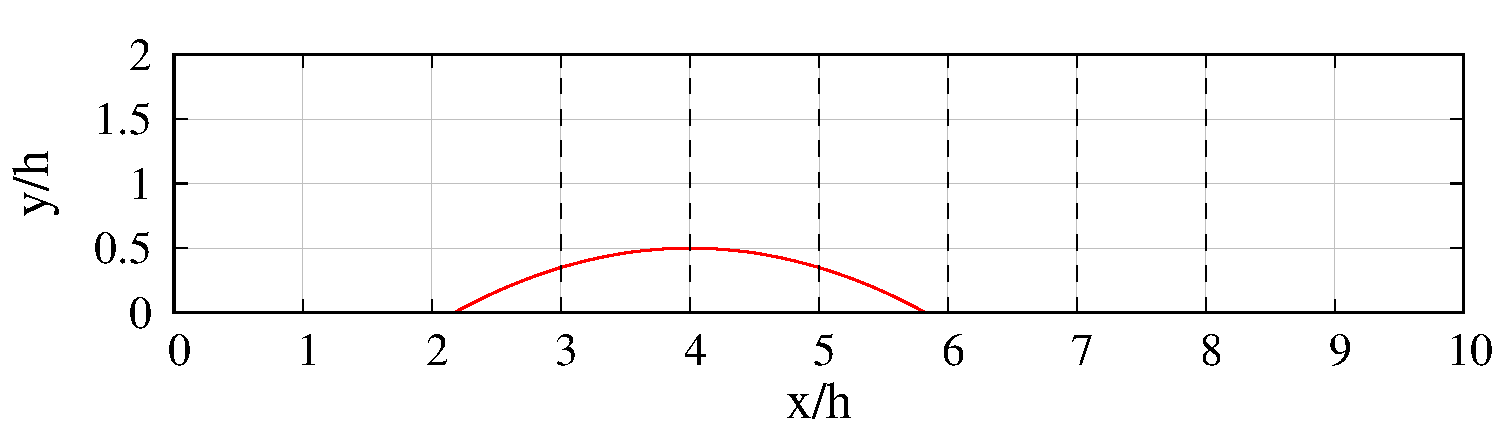}\mylab{-54.5mm}{22mm}{(\aaa)}\mylab{-47.5mm}{22mm}{(\bbb)}\mylab{-40.5mm}{22mm}{(\ccc)}\mylab{-33mm}{22mm}{(\ddd)}\mylab{-26mm}{22mm}{(\eee)}\mylab{-18.5mm}{22mm}{(\fff)}
  \includegraphics[width=0.95\linewidth]{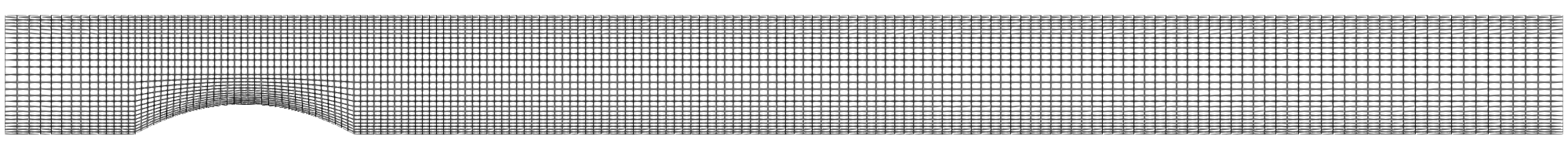}
  \includegraphics[width=0.95\linewidth]{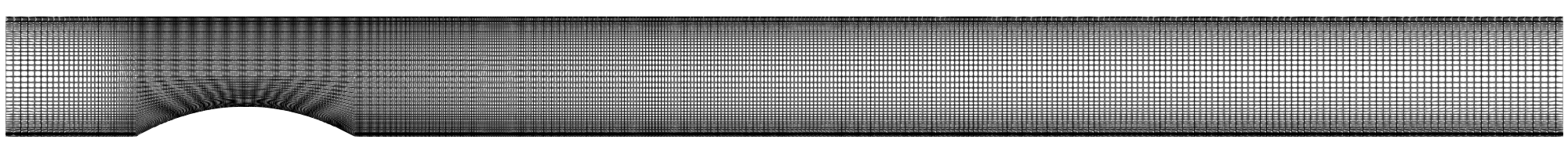}
  \caption{Geometry of the bump (top), numerical domain and mesh discretization for the coarse (central) and finer grid (bottom) of the turbulent flow over a bump. The dashed lines in the top panel report the streamwise positions (denoted with letters from {\itshape a} to {\itshape f})  where the mean and turbulent profiles are evaluated and reported in figures \ref{fig_bump_umean_M3} and \ref{fig_bump_uvar_M3}.}
\label{fig_bump_mesh}
\end{figure}

The numerical schemes and the settings of the dynamic RANS procedure are the same used for solving the previous flow configurations. Again, two grids have been generated to address the model's performance by varying the spatial resolution. The finer mesh is composed of $2.8M$ volumes that are stretched by moving towards the two walls in the wall-normal $y$-direction and by moving towards the bump center in the streamwise $x$-direction, see figure \ref{fig_bump_mesh}. The corresponding grid resolution is $\Delta x^+ = 15 \div 72$, $\Delta y_{wall}^+ = 2.5 \div 5$ and $\Delta z^+ = 30$. On the other hand, the coarser mesh is composed of $0.5M$ cells and the grid resolution is $\Delta x^+ = 60 \div 115$, $\Delta y_{wall}^+ = 10 \div 20$ and $\Delta z^+ = 30$. The wall-normal resolution for the coarser mesh has been intentionally designed to be coarse enough to address also the dynamic RANS performances when coupled with wall functions. Indeed, the T-Flows solver makes use of an elliptical blending function that, depending on the distance from the wall of the first cell, progressively activates the wall function required for very-coarse grained simulations of the near-wall region. In particular, the blending function for the coarse mesh leads approximately to 60\% of usage of the wall function among the whole domain while for the finer mesh to 0\% of usage meaning that no wall function is used.

\begin{figure}
  \centering
  \includegraphics[width=0.95\linewidth]{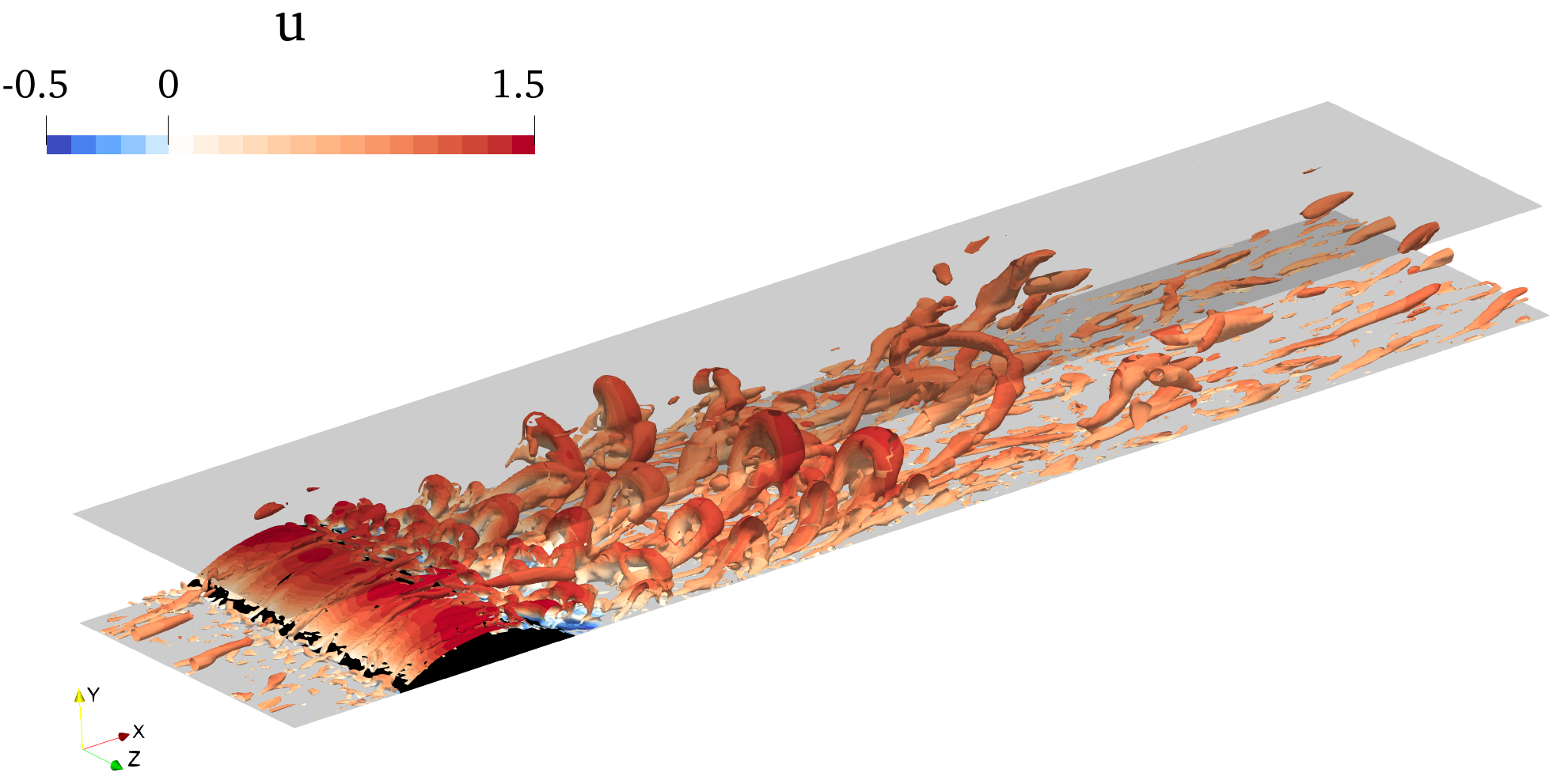}\vspace{-0.1cm}
  \includegraphics[width=0.95\linewidth]{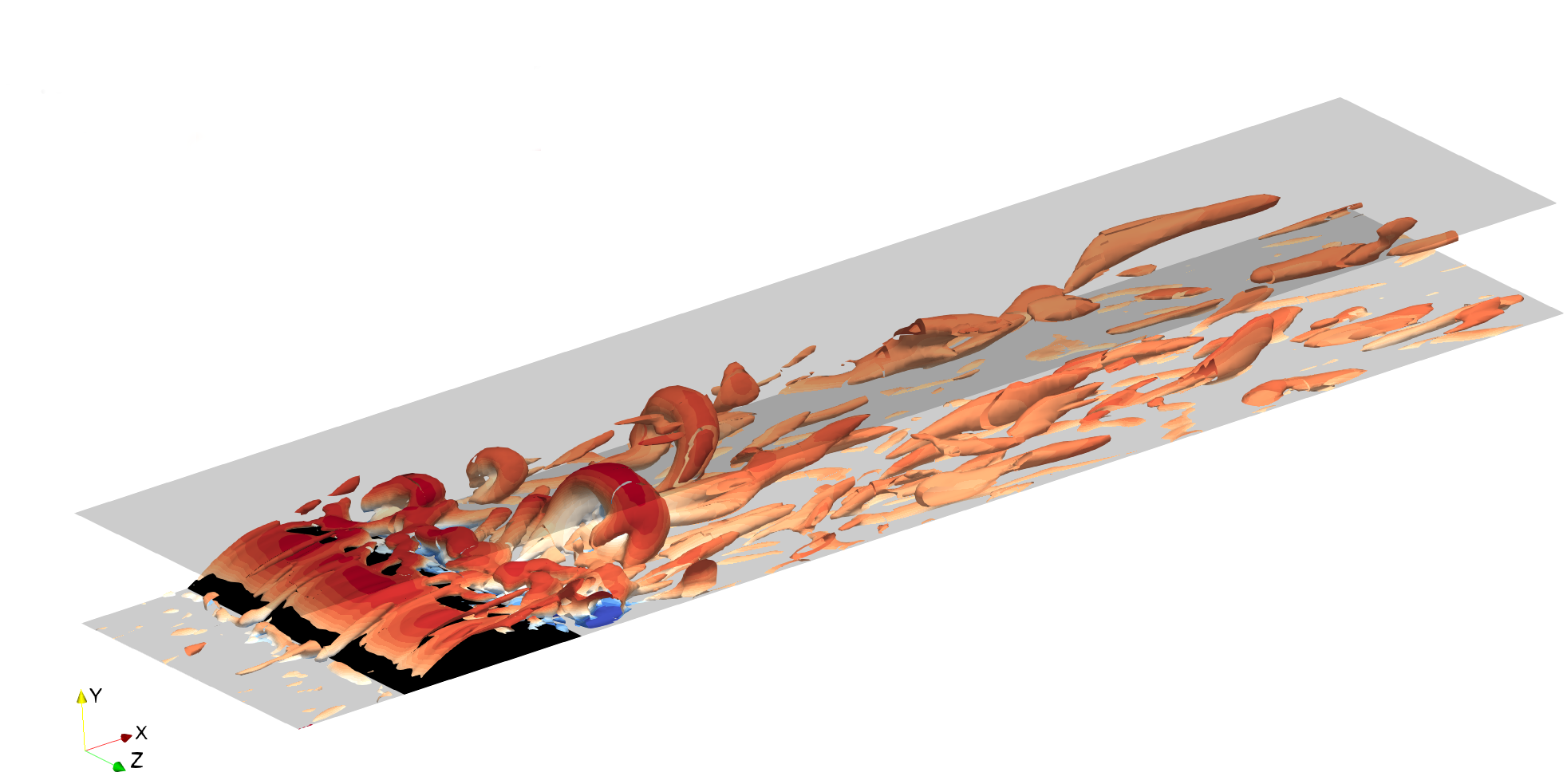}
  \caption{Turbulent flow over a bump. Instantaneous flow realizations reported with iso-surfaces the Q-Criterion \citep{hunt1988eddies}, $Q = 0.15$, colored with streamwise velocity. Solution of the dynamic RANS approach for the fine mesh (top) and for the coarse mesh (bottom).}
\label{fig_bump_inst_flow}
\end{figure}

We start the analysis of the results by addressing the instantaneous flow solution. As shown in figure \ref{fig_bump_inst_flow}, the dynamic RANS formalism is found again to provide many insights about the 3D and unsteady turbulent phenomena characterizing the flow solution. Although not reported for brevity reasons, the standard $k-\varepsilon$ implementation lacks in capturing this physics by reproducing a steady 2D flow pattern. Hence, the ability of the dynamic RANS approach in enabling scale-resolving features that are supported by the grid resolution is confirmed also in the fully turbulent conditions provided by the flow over a bump. The solution of the dynamic RANS formalism remains turbulent also when coarsening the grid with the only difference being the absence of fine-scale structures not supported by the spatial resolution, see again figure \ref{fig_bump_inst_flow}. Notice that in the coarse grid simulation wall functions are also widely adopted. Hence, the ability of the dynamic procedure in smoothly transitioning to a very-coarse modelling approach is further confirmed also by the flow over a bump.

\begin{table}
  \begin{center}
\def~{\hphantom{0}}
  \begin{tabular}{cccc}
                                     &  $x_s$ & $x_r$ & $c_d$ \\[3pt]
       DNS (Mollicone et al.)        &  4.4   &  6.2  & $2.05 \cdot 10^{-2}$ \\
       Dyn. $k-\varepsilon$ (fine)   &  4.4   &  6.6  & $2.04 \cdot 10^{-2}$ \\
       Dyn. $k-\varepsilon$ (coarse) &  4.6   &  7    & $1.84 \cdot 10^{-2}$\\
       $k-\varepsilon$ (fine)        &  4.9   &  6.2  & $1.93 \cdot 10^{-2}$\\
       $k-\varepsilon$ (coarse)      &  5.5   &  6.1  & $1.61 \cdot 10^{-2}$
  \end{tabular}
  \caption{Streamwise position of the mean flow separation $x_s$ and reattachment $x_r$ and drag coefficient $c_d$.}
  \label{tab_bump}
  \end{center}
\end{table}

\begin{figure*}
  \centering
  \includegraphics[width=0.9\linewidth]{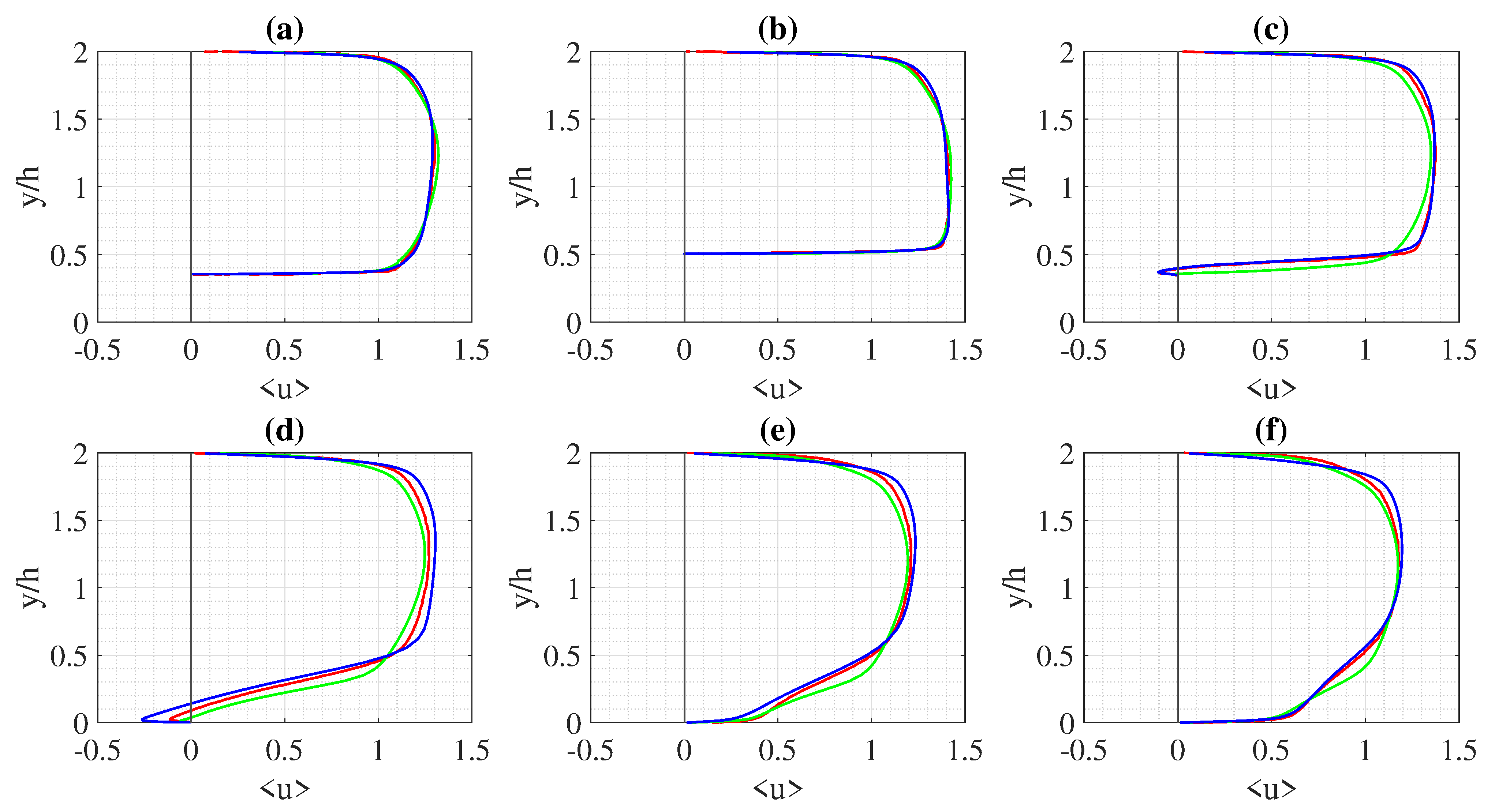}
  \caption{Turbulent flow over a bump. Mean streamwise velocity profiles at different streamwise positions (a), (b), (c), (d), (e) and (f) reported in figure \ref{fig_bump_mesh}. Finer grid solutions of the dynamic RANS (blue line) and standard $k-\varepsilon$ (green line) are compared with DNS data \citep{mollicone2017effect} (red line).}
\label{fig_bump_umean_M3}
\end{figure*}

From a more quantitative point of view, we analyse first the statistical features of the recirculating bubble generated by the bump. As shown in table \ref{tab_bump}, the average position of the flow separation $x_s$ is nicely captured by the dynamic procedure compared to DNS data \citep{mollicone2017effect} for both the fine and coarse simulations. On the contrary, the standard $k-\varepsilon$ approach significantly fails in predicting the phenomenon of separation especially in the coarse simulation. Contrary to the dynamic RANS formalism, the standard $k-\varepsilon$ model is not able to capture the damping of turbulence induced by the favorable pressure gradient in the windward side of the bump thus drastically delaying the point of separation. For the same reason, a shrinking of the size of the separation bubble is predicted by the standard $k-\varepsilon$ solution. Indeed, the reattachment length $x_r$ is only slightly overestimated by the dynamic RANS approach but not by the standard $k-\varepsilon$ solution, see again table \ref{tab_bump}. The behaviour of the flow separation and reattachment drastically influences the overall drag coefficient of the flow. As shown in table \ref{tab_bump}, the predicted drag coefficient is indeed improved by the dynamic RANS approach compared with the standard $k-\varepsilon$ solution.

The mean streamwise velocity profiles obtained with the fine grid simulations of the dynamic RANS and standard $k-\varepsilon$ approaches are shown in figure \ref{fig_bump_umean_M3} compared with those of the DNS by \citet{mollicone2017effect}. The profiles are reported for relevant streamwise flow positions starting from the windward side of the bump and ending in the reattached flow region as highlighted with letters from {\itshape a} to {\itshape f} in figure \ref{fig_bump_mesh}. The DNS behaviour is well reproduced by the dynamic RANS solution. Compared with the standard $k-\varepsilon$ approach, a significant improvement is observed especially at streamwise positions corresponding to the recirculating bubble and to the downstream flow development (streamwise positions {\itshape c-f}). This improvement supports the idea that the dynamic RANS approach outperforms the standard $k-\varepsilon$ model in predicting the turbulence weakening induced by favorable pressure gradients in the windward side of the bump with obvious repercussions for the downstream development of the solution. Roughly the same improvement is observed when coarsening the grid although these results are not reported here for the sake of brevity.

\begin{figure*}
  \centering
  \includegraphics[width=0.9\linewidth]{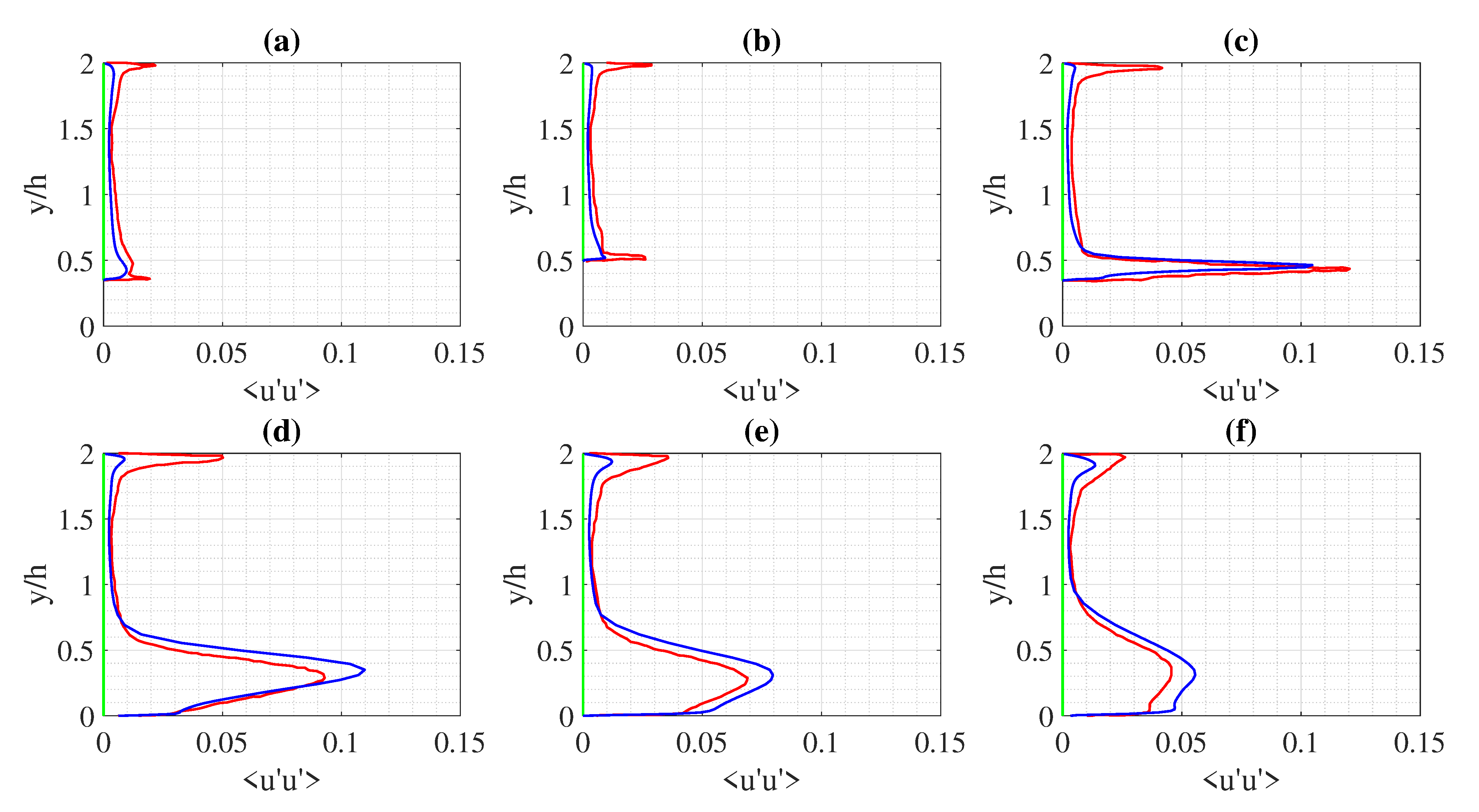}
  \caption{Turbulent flow over a bump. Streamwise velocity variance profiles at different streamwise positions (a), (b), (c), (d), (e) and (f) reported in figure \ref{fig_bump_mesh}. Finer grid solutions of the dynamic RANS (blue line) and standard $k-\varepsilon$ (green line) are compared with DNS data \citep{mollicone2017effect} (red line).}
\label{fig_bump_uvar_M3}
\end{figure*}

The streamwise velocity variance profiles obtained with the fine grid simulations of the dynamic RANS and standard $k-\varepsilon$ approaches are shown in figure \ref{fig_bump_uvar_M3}. The reference solution provided by the DNS of \citet{mollicone2017effect} is nicely captured by the dynamic RANS formalism. In particular, the shape of the streamwise velocity variance profiles, the positions of their peaks and also their intensity conform with those of the DNS thus further proving the quality of the dynamic RANS solution and of its capability in capturing the relevant physics of the flow. This physics is completely missed by the standard $k-\varepsilon$ approach that indeed reproduces a steady 2D solution.

\section{Effects of dynamic procedure settings}
\label{app_effects}

In this section, we address the robustness of the new modelling approach to the specific settings of the dynamic procedure detailed in section \S\ref{sec_dyn_settings} -- namely the correctness of removing the damping function $f_\mu$ from the definition of the eddy viscosity, and the small sensitivity of the solution to the width $T=N \Delta t$ of the test filter and to the upper bound of the clipping procedure for the model coefficient $c_\mu$. To this purpose, table \ref{tab:TIQ} compares the integral quantities and dominant frequency retrieved from a number of simulations of the BARC case on the coarser grid, with different values for the aforementioned setup parameters.

First, it is immediately clear that retaining $f_\mu$ leads to comparatively poor results, highlighting the improvement to near-wall behaviour brought by the dynamic procedure, as it eliminates the need for a damping function to compensate for the near-wall over-prediction of eddy viscosity predicted by the standard $k-\varepsilon$ model formulation. Indeed, using the damping function on top of the dynamic procedure causes excessive relaxation of the eddy viscosity, thus worsening the solution, as demonstrated in table \ref{tab:TIQ} by the overestimation of the integral quantities.

Concerning the width of the test filter $T=N \Delta t$, we found that the solution of the dynamic RANS approach is largely insensitive to its value. As shown in table \ref{tab:TIQ}, increasing the test filter width from $N=2$ to $N=3$ or $N=5$, the forces acting on the cylinder tend to slightly decrease --moving in the direction of the reference results--, but the difference is indeed marginal. The dominant frequency is also essentially unaltered.

Finally, concerning the effect of the upper limit ${c_\mu}_{max}$ for the clipping procedure of $c_\mu$, we found that the solution exhibits again a small sensitivity to its value. As shown in table \ref{tab:TIQ}, the Strouhal number of the dominant frequency and the integral forces are found to be almost unaltered when changing the upper limit of the clipping procedure from ${c_\mu}_{max}=0.2$ to ${c_\mu}_{max}=1$.

\begin{table}
	\begin{center}
		\def~{\hphantom{0}}
		\begin{tabular}{lccc}
			&  $c_d$ & $\sigma_{c_l}$ & $St$ \\[3pt]
			Ref. DNS             &  1.038 &  0.63 & 0.111\\
			Dyn. $k-\varepsilon$ (ref.) &  1.080 & 0.92 & 0.112\\[5pt]
			Dyn. $k-\varepsilon$ (w/ $f_\mu$) &  1.118 & 1.05 & 0.113\\[5pt]
			Dyn. $k-\varepsilon$ ($N=3$)           &  1.072 & 0.89 & 0.113\\
			Dyn. $k-\varepsilon$ ($N=5$)           &  1.072 & 0.88 & 0.113\\[5pt]
			Dyn. $k-\varepsilon$ (${c_\mu}_{max}=0.5$)     &  1.078 & 0.89 & 0.113\\
			Dyn. $k-\varepsilon$ (${c_\mu}_{max}=1.0$)     &  1.084 & 0.91 & 0.113
		\end{tabular}
		\caption{Summary of the mean integral quantities, $c_d$ and $\sigma_{c_l}$, and of the vortex shedding frequency $St$ of the simulations of the BARC case on the coarser grid, with and without the damping function $f_\mu$ for the eddy viscosity, and with different values of the width $T=N \Delta t$ of the test filter and of the upper bound of the clipping procedure for the model coefficient ${c_\mu}_{max}$.}
		\label{tab:TIQ}
	\end{center}
\end{table}


\section{Computational costs}
\label{app_comp_costs}

A careful analysis of the computational costs highlights that the improved quality of the flow solution with the dynamic RANS closure is not associated to an increase of the computational costs when considering a scale-resolving approach. In particular, the computational costs per time step across the different configurations and for each case of study for the 3D unsteady simulations, are essentially equivalent to those of the hybrid LES/RANS $k-\varepsilon-\zeta-f$ model. A different scenario appears when considering a statistical approach. In the 2D steady simulations, the computational costs per iteration are initially equivalent to those of the standard $k-\varepsilon$ model, but the latter eventually decrease by a factor of 4 as the solution stops changing among iterations. This reduction of computational costs is on the contrary not observed for the dynamic RANS approach given that its solution keeps changing among different iterations.

\bibliography{dynamic_rans}

@PREAMBLE{
 "\providecommand{\noopsort}[1]{}" 
 # "\providecommand{\singleletter}[1]{#1}%" 
}

@book{wilcox1998turbulence,
  title={Turbulence modeling for CFD},
  author={Wilcox, David C and others},
  volume={2},
  year={1998},
  publisher={DCW industries La Canada, CA}
}

@article{wilcox2008formulation,
  title={Formulation of the kw turbulence model revisited},
  author={Wilcox, David C},
  journal={AIAA journal},
  volume={46},
  number={11},
  pages={2823--2838},
  year={2008}
}

@inproceedings{spalart1992one,
  title={A one-equation turbulence model for aerodynamic flows},
  author={Spalart, Philippe and Allmaras, Steven},
  booktitle={30th aerospace sciences meeting and exhibit},
  pages={439},
  year={1992}
}

@article{menter1994two,
  title={Two-equation eddy-viscosity turbulence models for engineering applications},
  author={Menter, Florian R},
  journal={AIAA journal},
  volume={32},
  number={8},
  pages={1598--1605},
  year={1994}
}

@inproceedings{menter2005scale,
  title={A scale adaptive simulation model using two-equation models},
  author={Menter, Florian and Egorov, Yury},
  booktitle={43rd AIAA aerospace sciences meeting and exhibit},
  pages={1095},
  year={2005}
}

@article{FadaiGhotbi,
  title={Temporal filtering: A consistent formalism for seamless hybrid {RANS}--{LES} modeling in inhomogeneous turbulence},
  author={{Fadai-Ghotbi}, S. and {Friess}, C. and {Manceau}, R. and {Gatski}, T. and {Borée}, J.},
  journal={International Journal of Heat and Fluid Flow},
  volume={31},
  pages={378--389},
  year={2010}
}

@article{SSTDES,
  title			= "Ten Years of Industrial Experience with the {SST} Turbulence Model",
  author		= "Menter, F. R. and Kuntz, M. and Langtry, R.",
  journal		= "Turbulence, Heat and Mass Transfer",
  volume		= "4",
  pages			= "625--632",
  year			= "2003"
}

@article{leonard1975energy,
  title={Energy cascade in large-eddy simulations of turbulent fluid flows},
  author={{Leonard}, A.},
  journal={Adv. in geophysics},
  volume={18},
  pages={237--248},
  year={1975}
}

@article{klein2020analysis,
  title={Analysis and modelling of the commutation error},
  author={{Klein}, M. and {Germano}, M.},
  journal={Fluids},
  volume={6},
  number={1},
  pages={15},
  year={2020}
}

@article{moser2021statistical,
  title={Statistical properties of subgrid-scale turbulence models},
  author={Moser, Robert D and Haering, Sigfried W and Yalla, Gopal R},
  journal={Annual Review of Fluid Mechanics},
  volume={53},
  number={1},
  pages={255--286},
  year={2021},
  publisher={Annual Reviews}
}

@article{leonard2007commutation,
  title={Commutation error in LES with time-dependent filter width},
  author={Leonard, S and Terracol, M and Sagaut, P},
  journal={Computers \& fluids},
  volume={36},
  number={3},
  pages={513--519},
  year={2007},
  publisher={Elsevier}
}

@article{franke2001temporal,
  title={Temporal Commutation Errors in Large-Eddy Simulation},
  author={Franke, J and Frank, W},
  journal={ZAMM-Journal of Applied Mathematics and Mechanics/Zeitschrift f{\"u}r Angewandte Mathematik und Mechanik},
  volume={81},
  number={S3},
  pages={467--468},
  year={2001},
  publisher={Wiley Online Library}
}

@article{germano1992turbulence,
  title={Turbulence: the filtering approach},
  author={{Germano}, M.},
  journal={J. Fluid Mech.},
  volume={238},
  pages={325--336},
  year={1992}
}

@article{germano1991dynamic,
  title={A dynamic subgrid-scale eddy viscosity model},
  author={{Germano}, M. and {Piomelli}, U. and {Moin}, P. and {Cabot}, W.~H.},
  journal={Phys. Fluids},
  volume={3},
  number={7},
  pages={1760--1765},
  year={1991}
}

@article{lilly1992proposed,
  title={A proposed modification of the {G}ermano sugrid-scale closure method},
  author={{Lilly}, D.K.},
  journal={Phys. Fluids A},
  volume={4},
  pages={633--635},
  year={1992}
}

@article{meneveau2000scale,
  title={Scale-invariance and turbulence models for large-eddy simulation},
  author={{Meneveau}, C. and {Katz}, J.},
  journal={Annual Review of Fluid Mechanics},
  volume={32},
  number={1},
  pages={1--32},
  year={2000}
}

@article{lesieur1996new,
  title={New trends in large-eddy simulations of turbulence},
  author={Lesieur, Marcel and Metais, Olivier},
  journal={Annual review of fluid mechanics},
  volume={28},
  number={1},
  pages={45--82},
  year={1996}
}

@article{piomelli1999large,
  title={Large-eddy simulation: achievements and challenges},
  author={Piomelli, Ugo},
  journal={Progress in aerospace sciences},
  volume={35},
  number={4},
  pages={335--362},
  year={1999},
  publisher={Elsevier}
}

@incollection{launder1983numerical,
  title={The numerical computation of turbulent flows},
  author={{Launder}, B.E. and {Spalding}, D.B.},
  booktitle={Numerical prediction of flow, heat transfer, turbulence and combustion},
  pages={96--116},
  year={1983},
  publisher={Elsevier}
}

@article{yin2015dynamic,
  title={On the dynamic computation of the model constant in delayed detached eddy simulation},
  author={{Yin}, Z. and {Reddy}, K.~R. and {Durbin}, P.~A.},
  journal={Phys. Fluids},
  volume={27},
  number={2},
  year={2015}
}

@article{hanjalic2004robust,
  title={A robust near-wall elliptic-relaxation eddy-viscosity turbulence model for CFD},
  author={{Hanjali{\'c}}, K and {Popovac}, M and {Had{\v{z}}iabdi{\'c}}, M},
  journal={Int. J. Heat and Fluid Flow},
  volume={25},
  number={6},
  pages={1047--1051},
  year={2004}
}

@article{abe1995new,
  title={A new turbulence model for predicting fluid flow and heat transfer in separating and reattaching flows—II. Thermal field calculations},
  author={Abe, K and Kondoh, T and Nagano, Y},
  journal={International journal of heat and mass transfer},
  volume={38},
  number={8},
  pages={1467--1481},
  year={1995},
  publisher={Elsevier}
}

@article{hadvziabdic2022rational,
  title={A rational hybrid {RANS-LES} model for CFD predictions of microclimate and environmental quality in real urban structures},
  author={{Had{\v{z}}iabdi{\'c}}, M. and {Hafizovi{\'c}}, M. and {Ni{\v{c}}eno}, B. and {Hanjali{\'c}}, K.},
  journal={Building and Environment},
  volume={217},
  pages={109042},
  year={2022},
  publisher={Elsevier}
}

@article{delibra2010vortex,
  title={Vortex structures and heat transfer in a wall-bounded pin matrix: {LES} with a {RANS} wall-treatment},
  author={{Delibra}, G. and {Hanjali{\'c}}, K. and {Borello}, D. and {Rispoli}, F.},
  journal={Int. J. Heat Fluid Flow},
  volume={31},
  number={5},
  pages={740--753},
  year={2010}
}

@article{frohlich2008hybrid,
  title={Hybrid {LES/RANS} methods for the simulation of turbulent flows},
  author={{Fr{\"o}hlich}, J. and {Von Terzi}, D.},
  journal={Progress in Aerospace Sciences},
  volume={44},
  number={5},
  pages={349--377},
  year={2008}
}

@article{spalart2009detached,
  title={Detached-eddy simulation},
  author={Spalart, Philippe R},
  journal={Annual review of fluid mechanics},
  volume={41},
  number={1},
  pages={181--202},
  year={2009},
  publisher={Annual Reviews}
}

@article{heinz2020review,
  title={A review of hybrid RANS-LES methods for turbulent flows: Concepts and applications},
  author={Heinz, Stefan},
  journal={Progress in Aerospace Sciences},
  volume={114},
  pages={100597},
  year={2020},
  publisher={Elsevier}
}

@article{chaouat2017state,
  title={The state of the art of hybrid RANS/LES modeling for the simulation of turbulent flows},
  author={Chaouat, Bruno},
  journal={Flow, turbulence and combustion},
  volume={99},
  number={2},
  pages={279--327},
  year={2017},
  publisher={Springer}
}

@article{menter2021overview,
  title={An overview of hybrid RANS--LES models developed for industrial CFD},
  author={Menter, Florian and H{\"u}ppe, Andreas and Matyushenko, Alexey and Kolmogorov, Dmitry},
  journal={Applied Sciences},
  volume={11},
  number={6},
  pages={2459},
  year={2021},
  publisher={MDPI}
}

@incollection{spalart2021hybrid,
  title={Hybrid {RANS-LES} Methods},
  author={{Spalart}, P.~R.},
  booktitle={Advanced Approaches in Turbulence},
  pages={133--159},
  year={2021},
  publisher={Elsevier}
}

@inproceedings{kampe1951theoretical,
  title={Theoretical and experimental averages of turbulent functions},
  author={{Kamp{\'e} de F{\'e}riet}, J. and {Betchov}, R.},
  booktitle={Proc. K. Ned. Akad. Wet.},
  volume={53},
  pages={389--398},
  year={1951}
}

@article{pruett2003temporally,
  title={The temporally filtered Navier--Stokes equations: properties of the residual stress},
  author={{Pruett}, C.D. and {Gatski}, T.B. and {Grosch}, C.E. and {Thacker}, W.D.},
  journal={Phys. Fluids},
  volume={15},
  number={8},
  pages={2127--2140},
  year={2003}
}

@article{pruett2008temporal,
  title={Temporal large-eddy simulation: theory and implementation},
  author={{Pruett}, C.},
  journal={Theor. Comput. Fluid Dyn.},
  volume={22},
  pages={275--304},
  year={2008}
}

@article{fadai2010temporal,
  title={Temporal filtering: A consistent formalism for seamless hybrid {RANS--LES} modeling in inhomogeneous turbulence},
  author={{Fadai-Ghotbi}, A. and {Friess}, C. and {Manceau}, R. and {Gatski}, T.~B. and {Bor{\'e}e}, J.},
  journal={Int. J. Heat Fluid Flow},
  volume={31},
  number={3},
  pages={378--389},
  year={2010}
}

@article{friess2015toward,
  title={Toward an equivalence criterion for hybrid {RANS/LES} methods},
  author={{Friess}, C. and {Manceau}, R. and {Gatski}, T.~B.},
  journal={Computers \& Fluids},
  volume={122},
  pages={233--246},
  year={2015}
}

@article{duffal2022development,
  title={Development and validation of a new formulation of hybrid temporal large eddy simulation},
  author={{Duffal}, V. and {de Laage de Meux}, B. and {Manceau}, R.},
  journal={Flow, Turb. and Comb.},
  volume={108},
  number={1},
  pages={1--42},
  year={2022}
}

@article{hunt1988eddies,
  title={Eddies, streams, and convergence zones in turbulent flows},
  author={{Hunt}, J.~CR and {Wray}, A.~A. and {Moin}, P},
  journal={Center for Turbulence Research Report},
  volume={CTR-S88},
  pages={193},
  year={1988}
}

@article{cimarelli2024reynolds,
  title={Reynolds number effects in separating and reattaching flows with passive scalar transport},
  author={{Cimarelli}, A and {Corsini}, R and {Stalio}, E},
  journal={J. Fluid Mech.},
  volume={984},
  pages={A20},
  year={2024}
}

@article{cimarelli2018structure,
  title={On the structure of the self-sustaining cycle in separating and reattaching flows},
  author={{Cimarelli}, A. and {Leonforte}, A. and {Angeli}, D.},
  journal={J. Fluid Mech.},
  volume={857},
  pages={907--936},
  year={2018}
}

@Article{Crivellini20,
  author = 	 {F. Bassi and L. Botti and A. Colombo and A. Crivellini and M. Franciolini and A. Ghidoni and G. Noventa},
  title = 	 {A p-adaptive Matrix-Free Discontinuous Galerkin Method for the Implicit LES of Incompressible Transitional Flows},
  journal = 	 {Flow, Turbulence and Combustion},
  year = 	 {2020},
  volume = 	 {105},
  pages = 	 {437--470},
}

@article{yang2001large,
  title={Large-eddy simulation of boundary-layer separation and transition at a change of surface curvature},
  author={{Yang}, Z. and {Voke}, P.~R.},
  journal={J. Fluid Mech.},
  volume={439},
  pages={305--333},
  year={2001}
}

@Article{SMART,
  author = {P. H. Gaskell and A. K. C. Lau},
  title =  {Curvature-compensated convective transport: SMART, a new boundedness-preserving transport algorithm},
  year ={1988},
  journal = {International Journal for Numerical Methods in Fluids},
  volume = {8},
  pages ={617--641},
}

@article{cimarelli2020numerical,
  title={Numerical experiments in separating and reattaching flows},
  author={{Cimarelli}, A and {Franciolini}, M and {Crivellini}, A},
  journal={Phys. Fluids},
  volume={32},
  number={9},
  year={2020}
}

@Article{BARC_ChiariniQuadrio,
  author = {{Chiarini}, A. and {Quadrio}, M.},
  title =  {The turbulent flow over the BARC rectangular cylinder: A DNS study},
  year ={2021},
  journal = {Flow, Turbulence and Combustion},
  volume = {107},
  pages ={875--899},
}

@Article{BARC_Crivellini,
  author = {{Crivellini}, A. and {Nigro}, A. and {Colombo}, A. and {Ghidoni}, A. and {Noventa}, G. and {Cimarelli}, A. and {Corsini}, R.},
  title =  {Implicit Large Eddy Simulations of a rectangular 5:1 cylinder with a high-order discontinuous Galerkin method},
  year ={2022},
  journal = {Wind and Structures},
  volume = {34},
  pages ={59--72},
}

@article{Bruno,
  title={Benchmark on the aerodynamics of a rectangular 5:1 cylinder: an overview after the first four years of activity},
  author={{Bruno}, L. and {Salvetti}, M.~V. and {Ricciardelli}, F.},
  journal={J. Wind Eng. Ind. Aerodyn.},
  volume={126},
  pages={87--106},
  year={2014},
}

@Article{BARC_Mannini,
  author = {{Mannini}, C. and {Schewe}, G. and {Šoda}, A.},
  title =  {Unsteady RANS modelling of flow past a rectangular cylinder: Investigation of Reynolds number effects},
  year ={2010},
  journal = {Computers \& Fluids},
  volume = {39},
  pages ={1609--1624},
}

@article{wornom2011variational,
  title={Variational multiscale large-eddy simulations of the flow past a circular cylinder: Reynolds number effects},
  author={{Wornom}, S. and {Ouvrard}, H. and {Salvetti}, M.~V. and {Koobus}, B. and {Dervieux}, A.},
  journal={Computers \& Fluids},
  volume={47},
  number={1},
  pages={44--50},
  year={2011},
  publisher={Elsevier}
}

@article{mollicone2017effect,
  title={Effect of geometry and Reynolds number on the turbulent separated flow behind a bulge in a channel},
  author={{Mollicone}, J.-P. and {Battista}, F. and {Gualtieri}, P. and {Casciola}, C.~M.},
  journal={Journal of Fluid Mechanics},
  volume={823},
  pages={100--133},
  year={2017}
}

\end{document}